\documentclass[dvips]{arxstspdf}
\usepackage{graphicx}
\usepackage{flushend,stfloats}

\volume{25}
\issue{3}
\pubyear{2010}
\firstpage{388}
\lastpage{407}
\doi{10.1214/10-STS339}


\begin{document}
\begin{frontmatter}

\title{Laplace Approximated EM Microarray Analysis: An Empirical Bayes
Approach for Comparative Microarray Experiments}
\runtitle{Laplace Approximated EM Microarray Analysis}

\begin{aug}
\author[a]{\fnms{Haim} \snm{Bar}\ead[label=e1]{hyb2@cornell.edu}},
\author[b]{\fnms{James} \snm{Booth}\ead[label=e2]{jb383@cornell.edu}},
\author[c]{\fnms{Elizabeth} \snm{Schifano}\ead[label=e3]{eschifan@hsph.harvard.edu}}
\and
\author[d]{\fnms{Martin T.} \snm{Wells}\corref{}\ead[label=e4]{mtw1@cornell.edu}}
\runauthor{Bar, Booth, Schifano and Wells}

\affiliation{Cornell University, Cornell University, Harvard University
and Cornell University}

\address[a]{Haim Bar is Ph.D. candidate, Department of Statistical
Science, Cornell University, Ithaca, New York, USA (\printead{e1}).}
\address[b]{James G. Booth is Professor, Department of Biological
Statistics and Computational Biology, Cornell University, Ithaca, New York, USA (\printead{e2}).}
\address[c]{Elizabeth Schifano is Postdoctoral Fellow, Department of
Biostatistics, Harvard School of Public Health, Cambridge, Massachusetts, USA (\printead{e3}).}
\address[d]{Martin T. Wells is Professor, Department of Statistical
Science, Cornell University, Ithaca, New York, USA (\printead{e4}).}

\end{aug}

%
\begin{abstract}
A two-groups mixed-effects model for the comparison of (normalized)
microarray data from two treatment groups is considered. Most
competing parametric methods that have appeared in the literature
are obtained as special cases or by minor modification of the
proposed model. Approximate maximum likelihood fitting is
accomplished via a fast and scalable algorithm, which we call LEMMA
(Laplace approximated EM Microarray Analysis). The posterior odds of
treatment $\times$ gene interactions, derived from the model, involve
shrinkage estimates of both the interactions and of the gene
specific error variances. Genes are classified as being associated
with treatment based on the posterior odds and the local false
discovery rate (f.d.r.) with a fixed cutoff. Our model-based approach
also allows one to declare the non-null status of a gene by
controlling the false discovery rate (FDR). It is shown in a
detailed simulation study that the approach outperforms well-known
competitors. We also apply the proposed methodology to two previously
analyzed microarray examples.
Extensions of the proposed method to paired treatments and multiple
treatments are also discussed.
\end{abstract}

%

\end{frontmatter}

\section{Introduction}\label{sec1}
Microarray technologies have become a major data generator in the
post-genomics era. Instead of working on a gene-by-gene basis,
microarray technologies allow scientists\ to view the expression of
thousands of genes from an experimental sample simultaneously. Due
to the cost, it is common that thousands of genes are measured with
a small number of replications, as a consequence, one faces a \textit{
large G}, \textit{small n} problem, where $G$ is the total number of genes and
$n$ is the number of replications. After preprocessing of the raw
image data, the expression levels are often assumed to follow a
two-groups model, that is, the expressions are each either \textit{null}
or \textit{non-null} with prior probability $p_0 $ or $p_1 = 1 - p_0$,
respectively. The two-groups model plays an important role in the
Bayesian microarray literature and is broadly applicable (Efron, \citeyear{efron2008a}).

A general review of issues pertaining to microarray data analysis is
provided in Allison et al. (\citeyear{allicuipagesabr2006}). Here, we focus on
statistical inference and, in particular, on what
Allison et al. (\citeyear{allicuipagesabr2006}) refer to as ``consensus points 2 and
3'': the advantages of shrinkage methods, and controlling the false
discovery rate.\ We review several inferential methods, and develop a
unifying linear model approach.

Classical parametric statistics do not provide a \mbox{reliable} methodology
for determining differentially expressed genes. The large number of
genes with relatively few replications in typical microarray
experiments yield variance estimates of the expression levels that are
often unreliable. The classical $t$-test and $F$-test are generated under
a heterogeneous error variance model assumption and do not enjoy the
advantage gained by shrinkage estimation. The assumption that the
variances are equal across all genes is typically not realistic.
Hypothesis tests based on a pooled common variance estimator for all
genes have low power and can result in misleading differential
expression results (Wright and Simon, \citeyear{wrigsimo2003}; Smyth, \citeyear{smyt2004}; Cui et al., \citeyear{cuietal2005}).

An important observation is that, although there are only a few
replications for each gene, the total number of measurements is very large.
If information is combined across the genes (i.e.,
genome-wide shrinkage), it is possible to construct test procedures that
have improved performance. The SAM test (Tusher, Tibshirani and Chu, \citeyear{tusher2001}) and a
regularized $t$-test in Efron et al. (\citeyear{efron2001}) first used information
across the genome-wide expression values by the addition of a
data-based constant to the gene-specific standard errors.

The Bayesian approach seems to be particularly well suited for
combining information in expression data. Hierarchical Bayesian models
have also
been used for variance regularization by estimating moderated
variances of individual genes. The estimated variances are calculated
as weighted averages of the gene-specific sample variances and pooled
variances across all genes. In particular, the regularized $t$-test
proposed by Baldi and Long (\citeyear{baldilong}) uses a hierarchical model and
substitutes an empirical Bayes variance estimator based on a prior
distribution in place of the usual variance estimate.
Another hierarchical approach was developed in Newton et al.
(\citeyear{newtetal2001})
for detecting changes of gene expression in a two-channel cDNA
microarray experiment. This was extended to replicate chips with
multiple conditions using a hierarchical lognormal--normal model in
Kendziorski et al. (\citeyear{kendetal2003}). A key difference between these models and
those discussed above is that they effectively induce shrinkage in
the mean effects (i.e., the numerator of the $t$-statistic), while
assuming homogeneous variability across genes.

Instead of directly modeling the variation of the expression data,
two-groups models are characterized by mixing
measurements over latent gene-specific indicators.
Lonnstedt and Speed (\citeyear{lonnspee2002}) used this approach to
derive the so-called $B$-statistic as the logarithm of the posterior
odds of differential expression. Smyth (\citeyear{smyt2004}) extended the
$B$-statistic to the linear models setting and has written the widely
used \texttt{limma R} package (R Development Core Team, \citeyear{R2007}). Smyth (\citeyear{smyt2004}) also
shows that the $B$-statistic is a monotone function of a $t$-statistic
with a regularized variance which he refers to as a moderated
$t$-statistic. Wright and Simon (\citeyear{wrigsimo2003}) and Cui et al. (\citeyear{cuietal2005}) derive
similarly moderated statistics, and Cui et al. (\citeyear{cuietal2005}) showed that
their proposed test, using a James--Stein type variance estimator,
had the best or nearly the best power for detecting differentially
expressed genes over a wide range of situations compared to a number
of existing alternative procedures.

Since the performance of the $F$-type test statistics arising from
models with a random gene-specific error variance (leading to
shrinkage estimates of the error variances) is better than in the case
where the variances are fixed, why only model the variances as random
but not the means? In effect, the approach of Lonnstedt and Speed (\citeyear{lonnspee2002}),
and its extension in Smyth (\citeyear{smyt2004}), already do this by treating both
the gene-specific mean effects and error variances as random. These
models have been further generalized by
Tai and Speed (\citeyear{taispee2006,taispee2009}) to the multivariate setting to
handle, for example, short time-series of microarrays. These authors
coined the term ``fully moderated'' for such models. However, as we
point out later is Section \ref{sec:infer}, the specific
distributional assumptions made in these models imply that the
shrinkage factor for the mean effects is the same for all genes,
resulting in performance equivalent to the ordinary moderated-$t$.

\begin{table*}[b]

\caption{Models corresponding to combinations of fixed and random
factors}\label{inf.methods.table}
\begin{tabular*}{13cm}{@{\extracolsep{\fill}}lll@{}}
\hline
\multicolumn{1}{@{}l}{\textbf{Mean effect}} & \multicolumn{1}{c}{\textbf{Error variance}} & \multicolumn{1}{c@{}}{\textbf{Methods}} \\
\hline
\textbf{F}ixed & \textbf{F}ixed (Heterogenous) &
$t$-test/$F$-test \\
\textbf{F}ixed & Fixed (\textbf{H}omogenous) &
$F_3$ in Cui and Churchill (\citeyear{cuichur2003}) \\
\textbf{F}ixed & \textbf{R}andom & Wright and Simon (\citeyear{wrigsimo2003}), \\
& & Cui et al. (\citeyear{cuietal2005}),  \\
& & Lonnstedt and Speed (\citeyear{lonnspee2002}),\\
& & Smyth (\citeyear{smyt2004}) \\
\textbf{R}andom & \textbf{F}ixed (Heterogenous) & \\
\textbf{R}andom & Fixed (\textbf{H}omogenous) & Newton et al. (\citeyear{newtetal2001}),\\
& & Kendziorski et al. (\citeyear{kendetal2003}) \\
\textbf{R}andom & \textbf{R}andom & $F_{\mathrm{SS}}$ in Hwang and Liu (\citeyear{hwanliu2008}),  \\
& & Lonnstedt, Rimini and Nilsson (\citeyear{lonnrimi2005}),\\
& & Tai and Speed (\citeyear{taispee2009}),\\ & &
Lonnstedt and Speed (\citeyear{lonnspee2002}),\\
& & Smyth (\citeyear{smyt2004}) \\
\hline
\end{tabular*}
\end{table*}

Hwang and Liu (\citeyear{hwanliu2008}) proposed an alternative empirical Bayes approach
which shrinks both the means and variances differentially (see also
Liu, \citeyear{liu2006}). Their
simulation studies indicate that their fully moderated procedure is
more powerful than all the other tests existing in the
literature. The Hwang and Liu (\citeyear{hwanliu2008}) procedure uses method of moments
estimators of some model parameters rather than maximum likelihood. The
advantage of our EM fitting algorithm is that it is easily extended to
more general models, for example, including covariates, or the three
groups mixture model discussed in Section \ref{sec:modelext}.
Still, their
approach provided the key insight that
motivated the model formulation and subsequent computational algorithm
described in this article.

The development of the empirical Bayes methodologies that improve the
power to detect differentially expressed genes essentially reduces
to the choice of whether gene-specific effects should be
modeled as fixed or random. This question applies to effects on both
the mean and the error variance. Thus, there are four combinations of
fixed and random factors leading to four models which we denote by FF,
RF, FR and RR, where the first letter identifies whether the mean
effects are fixed or random and the second letter does the same for
the error variances. Two additional models, denoted FH and RH, are
obtained if the error variances are assumed to be homogeneous across
genes. The FF category corresponds to the naive approach of applying
$t$- or $F$-tests to each gene separately. The FR category includes the
models in Wright and Simon (\citeyear{wrigsimo2003}) and Cui et al. (\citeyear{cuietal2005}). The
gamma--gamma and log-normal--normal models of Newton et al. (\citeyear{newtetal2001}) and
Kendziorski et al. (\citeyear{kendetal2003}) are of the RH type. The approach of
Hwang and Liu (\citeyear{hwanliu2008}) falls in the RR category. Table \ref
{inf.methods.table} summarizes
how previously proposed statistics fall into the six model
categories.
Note that the RR category also includes the LIMMA model. However,
inference with the $B$-statistic of Lonnstedt and Speed (\citeyear{lonnspee2002}) and
Smyth (\citeyear{smyt2004}) results in a shrinkage factor for the mean effects which is
the same for all genes. Consequently, LIMMA is therefore similar to an
FR-type model in terms of frequentist performance since the posterior
odds are monotone in the moderated $t$-statistic.

In this paper we present a unified modeling framework for empirical
Bayes inference in microarray experiments together with a simple and
fast EM algorithm for estimation of the model parameters.
We focus on a simple two-condition experimental
setup, but the ANOVA formulation we posit in the next section allows
for easy generalization to more than two conditions and comparisons
based on a single sample of two channel arrays such as the more
general designs in Kerr, Martin and Churchill (\citeyear{kerrmartchur2000}) and Smyth (\citeyear{smyt2004}).
The methods of this article can, in principle, also be extended to a
multivariate empirical Bayes model, for example, to analyze
short time-course data as in the extension of the $B$-statistic by
Tai and Speed (\citeyear{taispee2006,taispee2009}), or to multiple array
platforms as is used in epigenomic data analysis
(Figueroa et al., \citeyear{figueroa2008}).

We apply an approximate EM algorithm for fitting the proposed model,
with the latent null/non-null status of each gene playing the role of
missing data. The integral needed to evaluate the complete data
likelihood makes direct application of the EM algorithm
intractable. However, a simple and accurate approximation is obtained
via the Laplace approximation (de~Bruijn, \citeyear{debr1981}, Chapter
4; Butler, \citeyear{butl2007}, page~42). This approximation makes the EM
algorithm scalable, tractable, and extremely fast. Implementation of
Bayesian microarray models typically involves drawing MCMC samples from
the posterior distribution of effects from all genes. MCMC sampling
provides a mechanism to study the full Bayesian posterior
distribution. However, there is a heavy computational burden that
makes the MCMC implementation less attractive. The Laplace
approximation circumvents the generation of the thousands of gene
effect parameters and gives a highly accurate approximation to the
integral in the expression of the complete data likelihood. The
Laplace approximated EM algorithm based analysis is the inspiration of
the acronym LEMMA (\textbf{L}aplace approximated \textbf{EM} \textbf{M}icroarray\textbf{A}nalysis) for the contributed R package, \texttt{lemma}
(Bar and Schifano, \citeyear{lemma2009}), which implements the methodology
described in this paper.

The paper is organized as follows. In Section \ref{sec:notation} we introduce the
necessary notation for our two-groups model along with the prior
distribution specifications. Section \ref{sec:estimation} describes the approximate EM
algorithm for fitting the RR model. We also propose a generalization
of the LIMMA model and show how the EM algorithm is easily
modified to estimate its parameters, and we briefly discuss
extensions to multiple treatments and to a three-groups model.
In Section \ref{sec:infer} we show that the posterior probability that a gene is
non-null is a function of a fully-moderated (in the sense of Hwang and Liu, \citeyear
{hwanliu2008}) posterior $t$-statistic with shrinkage in both the
numerator and the denominator. We show that our RR framework
generalizes several other statistics, and describe two inferential
procedures, one based on the posterior probability that a gene is
non-null, and one which is based on the null distribution and the FDR
procedure. Section
\ref{sec5} gives results of a simulation study in which we compare the
performance of various methods to the ``Optimal Rule'' procedure based on full
knowledge of the true model and its parameters. Our proposed
methodology is applied to two well-known microarray examples: the ApoA1 data
(Callow et al., \citeyear{dataapo}) and the Colon Cancer data (Alon et al., \citeyear{datacolonalon}) in
Section \ref{sec6}. We conclude the article in Section \ref{sec7} with a discussion.

\section{Model and Notation}\label{sec:notation}
Let $y_{ijg}$ denote the response (e.g., log expression ratio) of gene
$g$, for subject (replicate) $j$, in treatment group $i=1,2$. We begin
with the linear model,
%
\begin{eqnarray}
\label{eq:linear.model}
y_{ijg}=\mu+\tau_i+\gamma_g+\psi_{ig}+\varepsilon_{ijg},
\end{eqnarray}
with a typical assumption concerning the errors being
%
\begin{eqnarray}
\label{eq:error.distribution}
\varepsilon_{ijg}\sim\mbox{i.i.d. } N(0,\sigma_{\varepsilon,g}^2)
\end{eqnarray}
for $j=1,\ldots,n_{ig}$, independently across genes and treatment
groups. We impose the identifiability constraints, $\tau_1+\tau_2=0$
and $\psi_{1g}+\psi_{2g}=0$ for all $g=1,\ldots,G$. Then
$\tau=\tau_1-\tau_2$ is the main effect of treatment, averaged across
genes, and $\psi_g=\psi_{1g}-\psi_{2g}$, $g=1,\ldots, G$, are the gene
specific treatment effects. Note that we do not assume that the mean
treatment effect is zero. While assuming $\tau=0$ is often reasonable
when performing differential gene expression analysis on large
microarray data sets, we find this to be not only an unnecessary
constraint, but also unrealistic in certain situations. For example,
when a data set consists mostly of genes that are \textit{known} to be
differentially expressed, or when comparing expression levels across
species (where ``treatment'' is interpreted as ``species''), there is no
reason to assume that the overall mean difference between the two
treatment groups is zero.

We further suppose that the genes fall into two groups, a \textit{null}
group in which $\psi_g\equiv0$ and a \textit{non-null} group in
which $\psi_g\not= 0$. The primary goal is to classify
genes as null or non-null based on the observed responses. A
probabilistic approach is to suppose that each gene has prior
probability $p_1$ of being non-null (and $p_0=1-p_1$ of being null) and
to use Bayes rule to determine the posterior probability given the
data; specifically,
%
\begin{equation}
\label{eq:posterior.probability}
p_{1,g}(y_g)=\frac{p_1f_{1,g}(y_g)}{p_0f_{0,g}(y_g)+p_1f_{1,g}(y_g)} ,
\end{equation}
where $f_{1,g}(y_g)$ is the probability density of the responses for gene
$g$ implied by the non-null model, and $f_{0,g}(y_g)$ is the corresponding
quantity if the gene is in the null group.

In practice, of course, the mixture probability and the parameters
that determine the null and non-null densities have to be
estimated. This estimation step depends upon additional assumptions,
if any, that are made about the distribution of the responses.
As noted in the \hyperref[sec1]{Introduction}, a basic
question is whether gene-specific effects should be modeled as fixed or
random, leading to the model categories we denote by FF, RF, FR and RR,
and two additional models, FH and RH, obtained when the error variances
are assumed to be homogeneous, that is, $\sigma^2_{\varepsilon,g}\equiv
\sigma^2_\varepsilon$.

The ANOVA model (\ref{eq:linear.model}) together with the
distributional assumption (\ref{eq:error.distribution}) allows us to
restrict attention to the sum and difference of gene-specific
treatment means, respectively, $s_g=\bar{y}_{1\cdot g}+\bar{y}_{2\cdot
g}$ and
$d_g=\bar{y}_{1\cdot g}-\bar{y}_{2\cdot g}$, and the gene-specific mean squared
errors,
\[
m_g=\sum_{i=1}^2\sum_{j=1}^{n_{ig}}(y_{ijg}-\bar{y}_{i\cdot g})^2/f_g ,
\]
where $f_g=n_{1g}+n_{2g}-2$. Notice that $s_g|g \sim
N( 2\mu+2\gamma_g,\sigma_g^2)$, where
$\sigma_g^2\equiv\sigma_{\varepsilon,g}^2(1/n_{1g}+1/n_{2g})$, and $|g$
denotes conditioning on any gene-specific random effects. It follows
that $s_g$ carries no information about the gene-specific treatment
effect $\psi_g$. For this reason, our estimation procedures use only
the marginal likelihood based on the data $(\{d_g\},\{m_g\})$. The
model (\ref{eq:linear.model}) together with assumption
(\ref{eq:error.distribution}) also implies that $d_g$ and $m_g$ are
conditionally independent, with $d_g|g\sim(1-b_g)N_0+b_gN_1$
independently of $m_g|g\sim\sigma_{\varepsilon,g}^2\chi^2_{f_g}/\break f_g$,
where $b_g$, $g=1,\ldots,G$, denotes independent\break  $\operatorname{Bernoulli}(p_1)$
latent indicators of non-null status for the $G$ genes, $N_0$~and $N_1$
denote normal variates with unequal means $\tau$ and
$\tau+\psi_g\not=\tau$ respectively, but equal variances $\sigma_g^2$,
and $\chi^2_{f_g}$ denotes a chi-squared variate with $f_g$ degrees of
freedom.

The family of parametric models considered in this paper is completed
by specifying distributions for the gene-specific effects,
$\{\psi_g\}$ and $\{\sigma_{\varepsilon,g}^2\}$. In what follows we
suppose that, if the (non-null) gene-specific effects are
modeled as random variates, they follow a normal distribution,
%
\begin{equation}
\label{eq:psi.distribution}
\psi_g \sim \mbox{i.i.d. } N(\psi,\sigma_{\psi}^2).
\end{equation}
On the other hand, if the gene-specific variances are modeled as
random variates, they are drawn from an inverse gamma distribution,
%
\begin{equation}
\label{eq:sigma.distribution}
\sigma_{\varepsilon,g}^{-2}\sim \mbox{i.i.d.} \operatorname{Gamma}(\alpha,\beta) ,
\end{equation}
where $\alpha$ and $\beta$ are shape and scale parameters. We refer to
the RR model specified by (\ref{eq:linear.model}),
(\ref{eq:error.distribution}) and (\ref{eq:sigma.distribution}) with
the non-null gene-specific effects (\ref{eq:psi.distribution}) as the\break
LEMMA model.

It is worth contrasting (\ref{eq:psi.distribution}) with the
corresponding assumption in the models leading to the $B$-statistic
given in Lonnstedt and Speed (\citeyear{lonnspee2002}) and Smyth (\citeyear{smyt2004}), where the mean
of the random effects distribution is assumed to be zero. In a
classical (one group) normal mixed-model, the mean of the random
effect is assumed to be zero because it is not separately identifiable
from the overall mean. However, in the two-groups setting in which
$\psi_g$ in (\ref{eq:linear.model}) is modeled as a mixture, assuming
$\psi\neq0$ in (\ref{eq:psi.distribution}) poses no such
identifiability problems. Furthermore, this additional parameter
allows for two useful and important extensions of the model: (a)~to
paired (within-group) analyses, and (b)~to three-groups allowing for
over- and under-expressed non-null status. These extensions are
described in more detail in Section \ref{sec:modelext}.

\section{Estimation}
\label{sec:estimation}
In this section we describe in detail an approximate EM algorithm for
fitting the LEMMA model.\vadjust{\goodbreak} Estimation for the other five models can be
carried out by making appropriate modifications to this algorithm. The
LEMMA model has six parameters, two being the shape and scale of the
distribution for the error variances given in
(\ref{eq:sigma.distribution}). The remaining vector of parameters is
$( p_{1},\tau,\psi,\sigma_{\psi}^{2})$ which we denote by
$\phi$.

Estimates of the hyperparameters, $\alpha$ and $\beta$, are obtained
by maximizing the marginal likelihood based on $\{m_g\}$, given by
%
\begin{eqnarray}\label{eq:likelihood.ab}
&&L(\{m_g\}) \nonumber\\
&&\quad =  \prod_{g=1}^G\int_0^\infty
f(m_g|\sigma^2_{\varepsilon,g})f(\sigma^{-2}_{\varepsilon,g})\,d\sigma
^{-2}_{\varepsilon,g}
\nonumber
\\[-8pt]
\\[-8pt]
\nonumber
&&\quad=  \prod_{g=1}^G\frac{m_g^{{f_g}/{2}-1}
({f_g}/{2})^{{f_g}/{2}}}
{\Gamma({f_g}/{2})\Gamma(\alpha)\beta^\alpha}\\
&&\hspace*{38pt}{}\cdot\frac{\Gamma({f_g}/{2}+\alpha)}
{({m_g f_g}/{2} + {1}/{\beta} )^{{f_g}/{2}+\alpha}}.\nonumber
\end{eqnarray}
In practice, we find the maximum likelihood estimates for $\alpha$ and
$\beta$ using the {\tt nlminb} function in R. In all the simulations
and case studies the function converged quickly. Since the marginal
likelihood is based on the statistics $\{m_g\}$, the computation time
depends only on the number of genes, $G$, but not on the sample sizes.
We have also derived and implemented moment estimators [similar to
Smyth (\citeyear{smyt2004}), who comments that $\{m_g\}$ follow a scaled
$F$-distribution], and we found that both methods provide accurate
estimation of $\alpha$ and $\beta$.

\subsection{EM Algorithm}\label{sec:em}
\label{subsec:em}
We apply the EM algorithm to estimate $\phi$, with the
latent indicators, $\{b_{g}\}$, playing the role of the missing data.
Since $d_g$ and $m_g$ are conditionally\vspace*{1pt} independent given
$(b_g,\sigma_{\varepsilon,g}^2)$, the complete data likelihood for $\phi$
based on ($\{b_g\},\{d_g\},\{m_g\}$) is
%
\begin{eqnarray}
\label{eq:complete.likelihood}
L_C(\phi) &=& \prod_{g=1}^G \int L(b_g, d_g;
\sigma_{\varepsilon,g}^{2})
\nonumber
\\[-8pt]
\\[-8pt]
\nonumber
&&\hspace*{23pt}{}\cdot L(m_g; \sigma_{\varepsilon,g}^{2})f(\sigma_{\varepsilon,g}^{-2})\,d\sigma
_{\varepsilon,g}^{-2},
\end{eqnarray}
where $f(\sigma_{\varepsilon,g}^{-2})$ represents the gamma density with
shape $\alpha$ and scale $\beta$.

The integral in (\ref{eq:complete.likelihood}) makes direct
application of the EM algorithm intractable. However, a simple and
accurate approximation is obtained via the Laplace approximation
(de~Bruijn, \citeyear{debr1981}, Chapter 4; Butler, \citeyear{butl2007}, page 42)
%
\begin{eqnarray}
\label{eq:laplace}
L_C(\phi)\approx\tilde{L}_C(\phi)&\equiv&\prod_{g=1}^G L(b_g,d_g;\tilde
{\sigma}_{\varepsilon,g}^{2})L(m_g; \tilde{\sigma}_{\varepsilon
,g}^{2})
\nonumber
\\[-8pt]
\\[-8pt]
\nonumber
&&\quad\hspace*{8pt}\cdot f(\tilde{\sigma}_{\varepsilon,g}^{-2})\sqrt{-2\pi/\ell''(m_g;\tilde
{\sigma}^2_{\varepsilon,g})} ,
\end{eqnarray}
where $\ell''(m_g;\sigma^2_{\varepsilon,g})$ is the second derivative of\break
$\log L(m_g;\sigma^2_{\varepsilon,g})$ with respect to $\sigma^2_{\varepsilon
,g}$, and $\tilde{\sigma}_{\varepsilon,g}^2$ is the posterior mode of
$\sigma^2_{\varepsilon,g}$ given $m_g$,
given by
%
\begin{eqnarray}
\label{eq:posterior.mode}
\tilde{\sigma}_{\varepsilon,g}^2&=&\frac{f_{g}/2}{f_{g}/2+\alpha+1}m_{g}
\nonumber
\\[-8pt]
\\[-8pt]
\nonumber
&&{}+\frac{\alpha+1 }{f_{g}/2+\alpha+1}\cdot\frac{1}{(\alpha+1)\beta} .
\end{eqnarray}
Notice that the last three factors on the right-side of
(\ref{eq:laplace}) do not involve the parameter $\phi$ and can
therefore be ignored in the implementation of EM. In practice, we
replace $\alpha$ and
$\beta$ by their maximum likelihood estimates obtained from the
marginal likelihood in (\ref{eq:likelihood.ab}).

Denote the estimate after $m$ iterations of EM by $\phi^{(m)}$. The
$(m+1)$st E-step consists of taking the conditional expectation of the
logarithm of
(\ref{eq:complete.likelihood}) given the observed data, using the
current estimate, $\phi^{( m) }$. Using the Laplace
approximation (\ref{eq:laplace}), this is given by
\begin{eqnarray}\label{eq:q.function}
&&Q\bigl(\phi,\phi^{(m)}\bigr)\nonumber\\
&&\quad=E_{\phi^{(m)}}[\log L_C(\phi)|\{
d_g\}, \{m_g\}]
\nonumber\\
&&\quad\approx E_{\phi^{(m)}}[\log\tilde{L}_C(\phi)|\{d_g\}, \{m_g\}
]\nonumber\\
&&\quad=\sum_{g=1}^GE_{\phi^{(m)}}\{\log L(b_g,d_g;\tilde\sigma
_{\varepsilon,g}^2)|d_g\}+C\\
&&\quad=\sum_{g=1}^G\bigl\{p_{0,g}^{(m)}\log
[p_0f_{0,g}(d_g)]\nonumber\\
&&\qquad\hspace*{17pt}{}+p_{1,g}^{(m)}\log[p_1f_{1,g}(d_g)]\bigr\}+C
\nonumber\\
&&\quad\equiv \tilde{Q}\bigl(\phi,\phi^{(m)}\bigr)+C,\nonumber
\end{eqnarray}
where $C$ does not depend on $\phi$, $f_{0,g}$ and $f_{1,g}$ denote
$N(\tau,\tilde\sigma_g^2)$ and
$N(\tau+\psi,\sigma_\psi^2+\tilde\sigma_g^2)$ densities with
$\tilde\sigma_g^2=\tilde\sigma^2_{\varepsilon,g}(1/n_{1g} + 1/n_{2g})$,
$p_{1,g}=E(b_g|d_g)$ and $p_{0,g}+p_{1,g}=1$.

The M-step at the $(m+1)$ iteration requires maximization of $\tilde
{Q}(
\phi, \phi^{(m)} )$ with respect to $\phi$ to yield the
updated estimate $\phi^{(m+1)}$. That is,
\[
\phi^{( m+1) }=\arg\max_{\phi} \tilde{Q}\bigl( \phi, \phi
^{(m)} \bigr) .
\]
This leads to the following maximum likelihood estimate update
equations for $p_1$, $\tau$ and $\psi$:
%
\begin{eqnarray}
\label{eq:p1.update}
p_{1}^{( m+1) }&=&\frac{1}{G}\sum_{g=1}^{G}p_{1,g}^{(m)} ,
\\
\label{eq:tau.update}
\tau^{(m+1)}&=&
\frac{\sum_{g=1}^{G}p_{0,g}^{(m)}d_{g}/\tilde{\sigma}_g^{2}}
{\sum_{g=1}^{G}p_{0,g}^{(m)}/\tilde{\sigma}_g^2}
\end{eqnarray}
and
%
\begin{eqnarray}
\label{eq:psi.update}
\qquad&&\psi^{(m+1)}
\nonumber
\\[-8pt]
\\[-8pt]
\nonumber
&&\quad=
\frac{\sum_{g=1}^{G}p_{1,g}^{(m)}(d_g-\tau^{(m+1)})/
(\sigma_{\psi}^{2(m)}+\tilde{\sigma}_g^{2})}
{\sum_{g=1}^{G}p_{1,g}^{(m)}/(\sigma_\psi^{2(m)}+\tilde{\sigma
}_g^{2})} ,
\end{eqnarray}
while the update for $\sigma_\psi^{2}$ is the solution of the equation
%
\begin{eqnarray}
\label{eq:sigma2psi.update}
&&\sum_{g=1}^{G}p_{1,g}^{(m)}\frac{1}{\sigma_\psi^2+\tilde{\sigma}_g^2}
\nonumber
\\[-8pt]
\\[-8pt]
\nonumber
&&\quad=\sum_{g=1}^{G}p_{1,g}^{(m)}\frac{(d_g-\tau^{(m+1)}-\psi
^{(m+1)})^2}
{( \sigma_\psi^2+\tilde{\sigma}_g^2)^2} ,
\end{eqnarray}
and $\sigma_\psi^{2}=0$ if $p_{1,g}=0$ for all the genes.

Strictly speaking, the update for $\psi$ in (\ref{eq:psi.update}) is
conditional on the current value of $\sigma^2_\psi$. However, we have
found this variant of EM to have almost identical convergence
properties to the full EM in which $\tilde{Q}$ is maximized jointly with
respect to all four components of~$\phi$.

\subsection{Modifications for RF, RH, FF, FH, FR}\label{sec:mods}
LEMMA is considered an RR model because the gene-specific effects ($\psi
_g, \sigma^2_{\varepsilon,g}$) are modeled as random variates.
By considering one or both of these as fixed effects, we obtain models
that fall into one of the RF, RH, FR, FF or FH categories. Henceforth,
the category labels RF, RH, FF, FH, FR refer to the models derived from
the LEMMA (RR) model with the corresponding fixed/random distributional
assumption modifications.

The complete data likelihood for the RF model is
%
\begin{equation}
\label{eq:complete.likelihood.RF}
L_C(\phi)\approx \prod_{g=1}^G L(b_g, d_g; \sigma_{\varepsilon,g}^{2})L(m_g; \sigma
_{\varepsilon,g}^{2}) .
\end{equation}
Since no integration is required to evaluate this likelihood, the
Laplace approximation is not needed in this case. As with the LEMMA
(RR) model,
we first estimate the error variances, $\{\sigma_{\varepsilon,g}^2\}$,
separately using the marginal likelihood for $\{m_g\}$. This results
in the simple estimate, $\hat\sigma_{\varepsilon,g}^2=m_g$. The EM
algorithm for estimating $\phi$ then proceeds in an identical manner
except that $\tilde\sigma_g^2$ is replaced by
$\hat\sigma_g^2=\hat\sigma_{\varepsilon,g}^2(1/n_{1g}+1/n_{2g})$. The
algorithm for the RH model is also similar with the marginal
likelihood estimator of the homogeneous error variance given by
$\hat\sigma_\varepsilon^2=\sum_gm_gf_g/\sum_gf_g$.

For all the fixed gene-specific effects models (FR, FF and FH) it is easily
verified that $d_g-\tau^{(m)}-\psi_g^{(m)}=0$. This implies that the EM
update for the mixing parameter satisfies
\begin{eqnarray*}
&&p_1^{(m+1)}\\
&&\quad=\frac{1}{G}\sum_{g=1}^G\frac{p_1^{(m)}}
{p_0^{(m)}\exp\{-(d_g-\tau^{(m)})^2/2\hat{\sigma}^{2}_{e,g}
\}+p_1^{(m)}}\\
&&\quad>p_1^{(m)} ,
\end{eqnarray*}
where $\hat{\sigma}^{2}_{e,g}$ represents the appropriate $\sigma^2_g$
estimator for the desired model. As a result, the EM sequence for
$p_1$ always converges to 1, regardless of the starting value. An
explanation for this behavior is that the mixture probability is not
identifiable if the gene-specific effects are fixed.

\subsection{A Generalization of LIMMA}\label{sec:genlimma}
The LIMMA model proposed by Smyth (\citeyear{smyt2004}) is similar to the
LEMMA model described in Section~\ref{sec:notation}. A key difference
is the
assumption concerning the random gene-specific effects given in
(\ref{eq:psi.distribution}). The corresponding assumption in LIMMA is
$\psi_g|\sigma_{\varepsilon,g}^2\sim
N(0,v_0\sigma_{\varepsilon,g}^2)$. This assumption, combined with
(\ref{eq:sigma.distribution}), results in a closed form expression
for the complete data likelihood (\ref{eq:complete.likelihood}),
rendering the use of the Laplace approximation unnecessary. Another difference
is that the mean effect of treatment, averaged across genes ($\tau$),
is assumed to be zero in the LIMMA
model. However, this difference has little bearing on the arguments
that follow.

As noted in Section \ref{sec:notation}, it is
unnecessary to assume that the mean of the non-null gene-specific
effects, $\psi$, is zero. Hence, we consider a generalized LIMMA model
(denoted by RG in what follows) with
%
\begin{equation}
\label{eq:psi.LIMMA}
\psi_g|\sigma_{\varepsilon,g}^2\sim N(\psi,v_0\sigma_{\varepsilon,g}^2)
\end{equation}
for the non-null gene-specific effects, and, as such, it falls into the
RR category.
The EM algorithm discussed earlier in this section can be implemented
to fit this generalized model with minor modifications. Specifically,
after using the Laplace approximation, the $Q$-function has the same
form as (\ref{eq:q.function}) with
$v_{0,g}\tilde\sigma_{\varepsilon,g}^2$ replacing
$\sigma_\psi^2+\tilde\sigma_g^2$ as the variance in the non-null
density $f_{1,g}$, where
$v_{0,g}=v_0+1/n_{1g}+1/n_{2g}$. This leads to update
equations for $p_1$ and $\tau$ identical to (\ref{eq:p1.update}) and
(\ref{eq:tau.update}), respectively. The update for $\psi$ is
\[
\psi^{(m+1)}=
\frac{\sum_{g=1}^{G}p_{1,g}^{(m)}(d_g-\tau^{(m+1)})/(v_{0,g}\tilde
{\sigma}_{\varepsilon,g}^{2})}
{\sum_{g=1}^{G}p_{1,g}^{(m)}/(v_{0,g}\tilde{\sigma}_{\varepsilon
,g}^{2})} ,
\]
and the update of $v_0$ satisfies
\[
\sum_{g=1}^Gp_{1,g}^{(m)}\frac{1}{v_{0,g}}=\sum
_{g=1}^Gp_{1,g}^{(m)}\frac{(d_g-\tau^{(m+1)}-\psi
^{(m+1)})^2}{v_{0,g}^2\tilde\sigma_{\varepsilon,g}^2} ,
\]
and $v_0=0$ if $p_{1,g}=0$ for all the genes.
These updates simplify further if the sample sizes are the same for all genes.

\subsection{Model Extensions}\label{sec:modelext}
The LEMMA model is easily extended in a number of useful ways. First,
it enables within-group analysis which follows the same estimation
procedure by simply dropping the $i$ index and combining the terms $\mu
$ and $\tau$. We found this to be useful in practical applications,
when, for example, researchers wish to perform a paired-sample test.

Similarly, we can extend the model to have multiple treatment groups
and test different (user-defined) contrasts, as was done in Smyth (\citeyear{smyt2004}) for the LIMMA model. Mathematically, this generalization is
very simple, and, in practice, when dealing with a small or moderate
number of treatment groups, the estimation procedure poses no
significant computational challenges.
For example, we use the ($t-1$-dimensional vector) summary statistics
$\mathbf{d}_{g}=\mathbf{H\bar{Y}}_{g},$
where $\mathbf{H}$ is a contrast matrix (e.g., the Helmert matrix)
and
$\mathbf{\bar{Y}}_{g} = ( \bar{Y}_{1\cdot g},\bar{Y}_{2\cdot
g},\ldots,\bar{Y}_{t\cdot g}) ^{\prime}$.
Note that the $2 \times2$ Helmert matrix gives the $d_g$ and $s_g$
statistics for the one-treatment case [scaled by a factor of
$1/\sqrt{2}$]. Obtaining the estimates and test statistics in the
multiple treatment case is analogous to the derivations in (\ref
{sec:em}). See the \hyperref[sec:app]{Appendix} for details.

As noted in Zhang, Zhang and Wells (\citeyear{zhanetal2010}), it is often the case that the
probabilities of under- and over-expressed genes are not equal. The
assumption that the distribution of the non-null genes has a nonzero
mean ($\psi$) can be modified to allow for multiple non-null components
in the mixture distribution. For example, we might assume that each
gene is either in the null group ($\psi_g=0$) with probability $p_0$,
in one non-null component with probability $p_1$ with $\psi_g \sim
\mbox{i.i.d. } N(\psi,\sigma_{\psi}^2)$, or in a second non-null group
with probability $p_2$ with $\psi_g \sim \mbox{i.i.d. } N(-\psi,\sigma
_{\psi}^2)$, where $p_0 +p_1 +p_2 = 1$. The two-component model in the
previous sections is the special case in which $p_2=0$. The \texttt
{lemma} \texttt{R} package uses the three component mixture by default,
and we have found that, indeed, when there are two mixture components,
the EM algorithm converges to $\hat{p}_2=0$. Note that the \texttt{R}
implementation assumes that the means of the non-null groups are of the
same magnitude but opposite sign. This assumption can be relaxed, for
instance, by assuming only that $\psi_1 < 0 < \psi_2$.

\section{Inference}\label{sec:infer}

The posterior probability that gene $g$ is non-null is given by the
expression (\ref{eq:posterior.probability}). Its estimated value based
on the LEMMA model can be expressed as a function of the likelihood ratio
%
\begin{eqnarray}\label{eq:likelihood.ratio}
\frac{L_{0,g}}{L_{1,g}}&\equiv&\frac{\hat{f}_{0,g}}{\hat{f}_{1,g}}\nonumber\\
&=& {(2\pi\tilde\sigma_g^2)^{-1/2}\exp\{ -(d_g-\hat\tau
)^2/2\tilde\sigma_g^2\}}\nonumber\\
&&{}\big/
\bigl([2\pi(\hat\sigma_\psi^2+\tilde\sigma_g^2)]^{-1/2}\nonumber\\
&&\hspace*{8pt}{}\cdot\exp\{ -(d_g-\hat\tau-\hat\psi)^2/2(\hat\sigma_\psi^2+\tilde\sigma
_g^2) \}\bigr)\\
&=& \biggl(\frac{\tilde\sigma_g^2}{\hat\sigma_\psi^2+\tilde\sigma
_g^2}\biggr)^{-1/2}\nonumber\\
&&{}\cdot\exp\biggl\{ -\frac{1}{2}\frac{[\hat\lambda_g(d_g-\hat\tau)+(1-\hat
\lambda_g)\hat\psi]^2}
{\hat\lambda_g\tilde\sigma_g^2}+\frac{\hat\psi^2}{2\hat\sigma_\psi
^2}\biggr\}
\nonumber\\
&\propto& \biggl(\frac{\tilde\sigma_g^2}{\hat\sigma_\psi^2+\tilde\sigma
_g^2}\biggr)^{-1/2}
\exp\biggl\{ -\frac{1}{2}T_g^2\biggr\} ,\nonumber
\end{eqnarray}
with the constant of proportionality being $\exp(\hat\psi^2/\break 2\hat\sigma
_\psi^2)$,
where
\[
\lambda_g=\frac{1}{\sigma_g^2}\biggl(\frac{1}{\sigma_g^2}+\frac
{1}{\sigma_\psi^2}\biggr)^{-1}
=\frac{\sigma_\psi^2}{\sigma_\psi^2+\sigma_g^2} .
\]
The statistic $T_g$ is a posterior $t$-statistic, being the ratio of the
estimated posterior expectation of $\psi_g$ to its estimated posterior
standard deviation. Note that the LEMMA model induces three forms of
shrinkage in $T_g$.
The first two forms come from $\hat{\lambda}_g > 0$ in both the numerator
and the denominator. Third, $\tilde{\sigma}^2_g$, a function of the
posterior mode $\tilde{\sigma}^2_{\varepsilon,g}$, is itself a shrinkage
estimator as a weighted compromise between the usual error variance
estimator $m_g$ and the mode of the inverse gamma distribution
$[(\alpha+ 1)\beta]^{-1}$.

The likelihood ratio in (\ref{eq:likelihood.ratio}) has the same form
for the RF
and RH models with $\tilde\sigma_{\varepsilon,g}^2$ replaced by
$\hat\sigma_{\varepsilon,g}^2$ and $\hat\sigma_\varepsilon^2$, respectively
in $\tilde\sigma_g^2$. [Recall that
$\sigma_g^2=\sigma_{\varepsilon,g}^2(1/n_{1g}+1/n_{2g})$.] Test
statistics for the fixed mean effects models, FR, FF and FH, are
obtained as limits of $T_g$ as $\hat\lambda_g\to1$.

It is interesting to compare the likelihood ratio
(\ref{eq:likelihood.ratio}) with the corresponding statistic under the\break
LIMMA and RG model assumptions discussed in the previous section. For
these models $\sigma_\psi^2$ is replaced by
$v_0\tilde\sigma_{\varepsilon,g}^2$, and so the shrinkage coefficient becomes
\[
\lambda_g=\frac{v_0}{v_0+1/n_{1g}+1/n_{2g}} .
\]
In particular, if the sample sizes are the same for all genes, then the
amount of shrinkage is the same for all genes. Furthermore, if $\psi$
is set equal to zero, as it is in LIMMA, then $T_g$ is proportional
to the test-statistic for the FR model,
\[
T_g=\frac{d_g-\hat\tau}{\tilde\sigma_g} .
\]
This has the same form as the moderated $t$-statistic of
Smyth (\citeyear{smyt2004}) and Wright and Simon (\citeyear{wrigsimo2003}) except for the
subtraction of the average gene effect, $\tau$, in the
numerator and the use of the mode rather than the expected value of
the posterior distribution of $\sigma_{\varepsilon,g}^2$ given $m_g$ in
the denominator.

For inference, we compare the posterior null probability, $1-p_{1,g}$ in
(\ref{eq:posterior.probability}), with a local f.d.r. threshold to decide
whether a gene is in the non-null group. Alternatively, our model-based
approach also allows one to declare the non-null status of a gene by
controlling the false discovery rate (FDR), using the
Benjamini and Hochberg (\citeyear{benjaminihochberg}) (BH) procedure for any given level, $q^*$.
Specifically, using the theoretical null-gene distributions of $\{d_g\}
$, which are assumed to be $N(\hat\tau, \tilde\sigma^2_{g})$, we obtain the
$p$-values for the observed $\{d_g\}$. We denote the $p$-values by $\{P_g\}
$, and find the largest index $g'$ for which $P^F_{g'} \le q^* \times
g'/G$, where $\{P^F_g\}$ is the sorted list of $p$-values.
We declare all the genes with index smaller than or equal to
$g'$ (in the sorted list) as non-null, and the FDR theorem guarantees that
the expected false discovery rate is bounded by $q^*$.

\section{Simulation Study}\label{sec5}
In this section we assess the performance of several
estimation/testing procedures mentioned in this paper under two data
generation models, one according to the LEMMA model and the other
according to the LIMMA model. In practice, the correct model is
unknown, so our goal is to compare the power, accuracy, false
discovery rate and parameter estimation for different
true-model/procedure combinations. In what follows we use
the term ``procedure'' to define the combination of the model selected
for analysis (which may or may not be the true model) and the
estimation and inferential techniques derived from this model.

\subsection{Data Generation}
In both scenarios (LEMMA and LIMMA), we simulated $S=100$ data sets
according to
a mixture model with two groups, null and non-null. Each data set
consisted of $G=2000$
genes, of which $p_1G$ were non-null, and we used $p_1=0.01, 0.05, 0.1, 0.25$.
For each of the $S$ data sets we drew $G$ inverse gamma error variates
with shape $\alpha$
and scale $\beta$. By varying
$\alpha$ and $\beta$, we adjusted the amount of error variance
variability present in the data. The values of $\alpha,$ $\beta$,
$n_{1g}\equiv n_1$, and $n_{2g}\equiv n_2$ were chosen so that
$\operatorname{mean}(\sigma^2_{g})=1$. With $n_1=n_2=6$, we set $\alpha= 5$
and $\beta= 1/12$ for the ``low'' error variance variability; we set
$\alpha= 2.1$ and $\beta= 10/33$ for the ``high'' error variance
variability. Hence, the standard deviation (and also the coefficient
of variation, CV) of $\sigma^2_g$ for the former was $1/\sqrt{3}$, and
for the latter was $\sqrt{10}$.

In the LEMMA-generated data, we varied $\psi,$ so that $\psi\in
\{0, 1, 2, \ldots, 6\} \equiv\Psi$, and set
$\sigma_{\psi}^2 = 1$. In the LIMMA data generation setup, we used
$v_0 \in\{\frac{1}{6}, \frac{2}{6},\break  \ldots, \frac{8}{6}\}$
to generate the non-null
genes according to (\ref{eq:psi.LIMMA}). For both generation schemes
we set $\tau=0,$ as the LIMMA model does not involve $\tau$, and it is
only estimable under the random gene by treatment interaction effect
models (RR, RF, RH). We generated $y_{ijg}$ according to equations
(\ref{eq:linear.model}) and (\ref{eq:error.distribution}) with the
above parameter specifications, and computed $\{d_g\}$ and
$\{m_g\}$. While we only present results for a selection
of specific parameter value settings, numerous simulations were
performed with a variety of sample sizes $n_i$, $i=1,2$, non-null
probabilities $p_1$, and gene-specific treatment variances
$\sigma^2_\psi$. In addition, we also considered using the log-normal
distribution to generate the error variance $\sigma^2_{\varepsilon,g}$
rather than the inverse gamma distribution. We found the results
to be qualitatively insensitive to these different settings, and the
results presented
below portray an accurate summary of the performance of the methods.

\subsection{Data Analysis and Results}\label{sec:analysis}

\begin{figure*}

\includegraphics{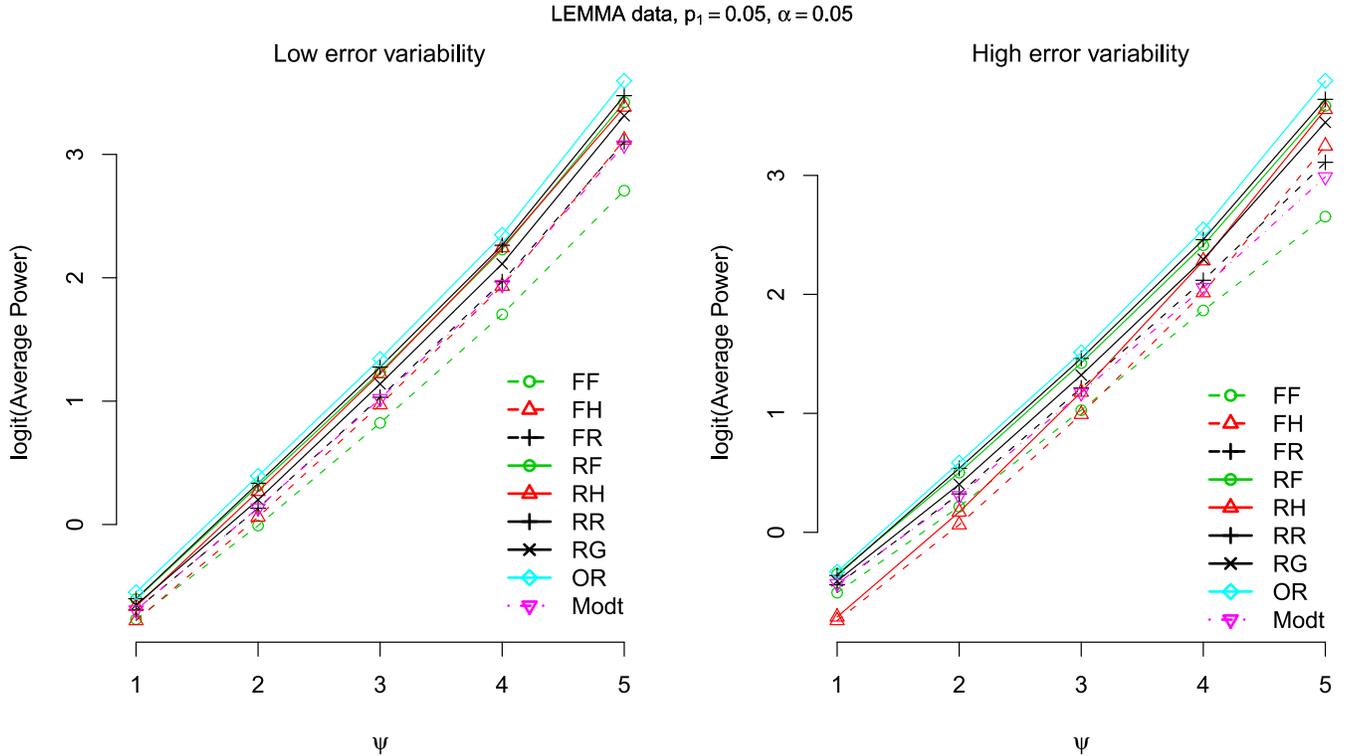}

\caption{Average power (on the logit scale) for empirical quantile
analysis under the RR data generation model, with
$n_1=n_2=6,$ $S=100$ samples, $G=2000$ genes, and $p_1=0.05$
probability of
non-null status. \textup{Left}: low error variance variability ($\mathit{CV}=0.58$).
\textup{Right}:
high error variance variability ($\mathit{CV}=3.16$).}
\label{fig:power.emp.quant}
\end{figure*}

We consider two metrics for determining null and non-null status of
genes. The first method is based on computing empirical quantile
critical values. Since the distribution of many of our test statistics
is unknown, we defined a test-specific critical value, $T_c$, as the
0.95 quantile among the $1900 \times100$ null genes. By design, this
resulted in an average size of 0.05 for each test. The average power
for each procedure was determined by the proportion of non-null genes
correctly declared non-null based on the (test-specific) empirical
critical value $T_c$. Figure \ref{fig:power.emp.quant} shows the
average power (on the logit scale) of the likelihood ratio tests
derived assuming the FF, FH, FR, RF, RH and RR models, with
estimation procedures as described in Section \ref{sec:estimation}.
Also included in our comparison were the RG likelihood ratio tests,
derived from the model defined in Section \ref{sec:genlimma}, and the
moderated $t$-tests obtained from the \texttt{limma R} package. Since
in our simulations we know the exact values of the parameters, we also
included the ``Optimal Rule'' statistics (denoted by OR) which were
obtained by plugging in the true parameter values in the likelihood
ratio statistic for the true data generation model (either LEMMA or
LIMMA).

When the data are generated according to the\break LEMMA model our
simulations show that the tests derived from the RR model achieved the
highest power for all $\psi\in\Psi$ (and almost identical to the
Optimal Rule's), as can be seen in Figure \ref{fig:power.emp.quant}.
When the data are generated according to the LIMMA model, the
likelihood ratio tests derived from the RR and RG models have nearly
identical performance in terms of power as those of the moderated-$t$
statistics and the LIMMA Optimal Rule for all values of $v_0$ (figure
not shown).

As expected, our simulations also showed that the average power in the
homogeneous error variance models (RH, FH) decreases as the error
variance variability increases. In general, the random gene models (RR,
RF, RH) demonstrate higher average power than their corresponding
fixed gene counterparts. Notice also that the performance of moderated-$t$ and the FR statistics are almost identical.

The second performance assessment method did not require computing
empirical quantiles, and
was based on local f.d.r. criteria. Efron et al. (\citeyear{efron2001}) and Efron (\citeyear
{efron2005}) defined local f.d.r. as
%
\begin{equation}
\operatorname{f.d.r.}(y_g) = \Pr(\operatorname{null} | Y=y_g)
\end{equation}
for the posterior probability of a gene $g$ being in the null group.
Note that this is precisely $1-p_{1,g}(y_g),$ where $p_{1,g}(y_g)$ is
given by (\ref{eq:posterior.probability}).
Since $p_1$ can only be estimated in the random-mean models, we only
considered the local f.d.r. statistics associated with RR, RF and RH. For
comparison, we also considered the local f.d.r. statistics for RG and the
Optimal Rule, and two types of $B$
statistics computed by the \texttt{limma} package
to differentiate between those
computed with the default
value of $p_1=0.01$ [referred to as ``Limma(0.01)''] and those
computed with the estimated value of $p_1$
[referred to as ``$\operatorname{Limma}(\hat{p}_1)$'']. We also included local f.d.r.
statistics computed from the \texttt{locfdr} (Efron, Turnbull and Narasimhan, \citeyear{Rlocfdr}) \texttt{R}
package (referred to as ``Efron'' for simplicity).

\begin{figure*}

\includegraphics{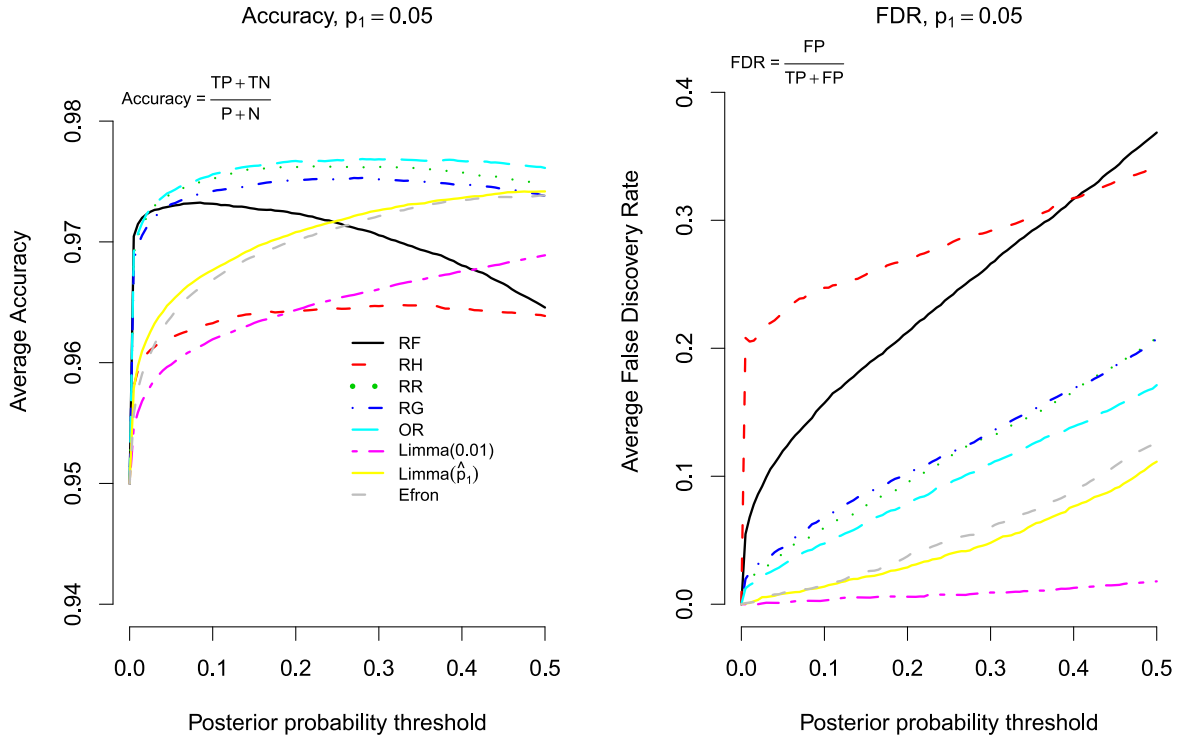}

\caption{Accuracy \textup{(left)} and false discovery rate \textup{(right)} for data
generated under the LEMMA model with
$n_1=n_2=6,$ $S=100$ samples, $G=2000$ genes, and $\psi=3$, $p_1=0.05$
probability of
non-null status, and high error variance variability ($\mathit{CV}=3.16$).}
\label{fig:RR.acc.fdr.1}
\end{figure*}

To evaluate the performance of these procedures, we looked at two
complementary metrics. The first is the measure of
accuracy, defined by the ratio $(\mathit{TP} + \mathit{TN}) / (P + N)$ as in
Hong (\citeyear{hong2009}), where $P$ and $N$ are the total numbers of non-null
and null genes, respectively, and $\mathit{TP}+\mathit{TN}$ is the sum of correct
classifications (true positives plus true negatives). The second
metric is the false discovery rate, defined by $\mathit{FP}/(\mathit{FP}+\mathit{TP})$,
where FP is the total number of false positives. Clearly,
our goal is to maximize the accuracy while maintaining a low false
discovery rate. To compare different methods, we computed the accuracy
and FDR for a range of posterior null probability thresholds (between
0 and 0.5). A gene is declared as non-null if its posterior null
probability is below the selected threshold. Note that when the
threshold is 0, all genes are declared as null and we obtain accuracy
of $1-p_1$. As we increase the threshold, the total number of
detections increases, and if we let the threshold be 1, all genes are
declared as non-null (and the accuracy is $p_1$).

\begin{figure*}[b]

\includegraphics{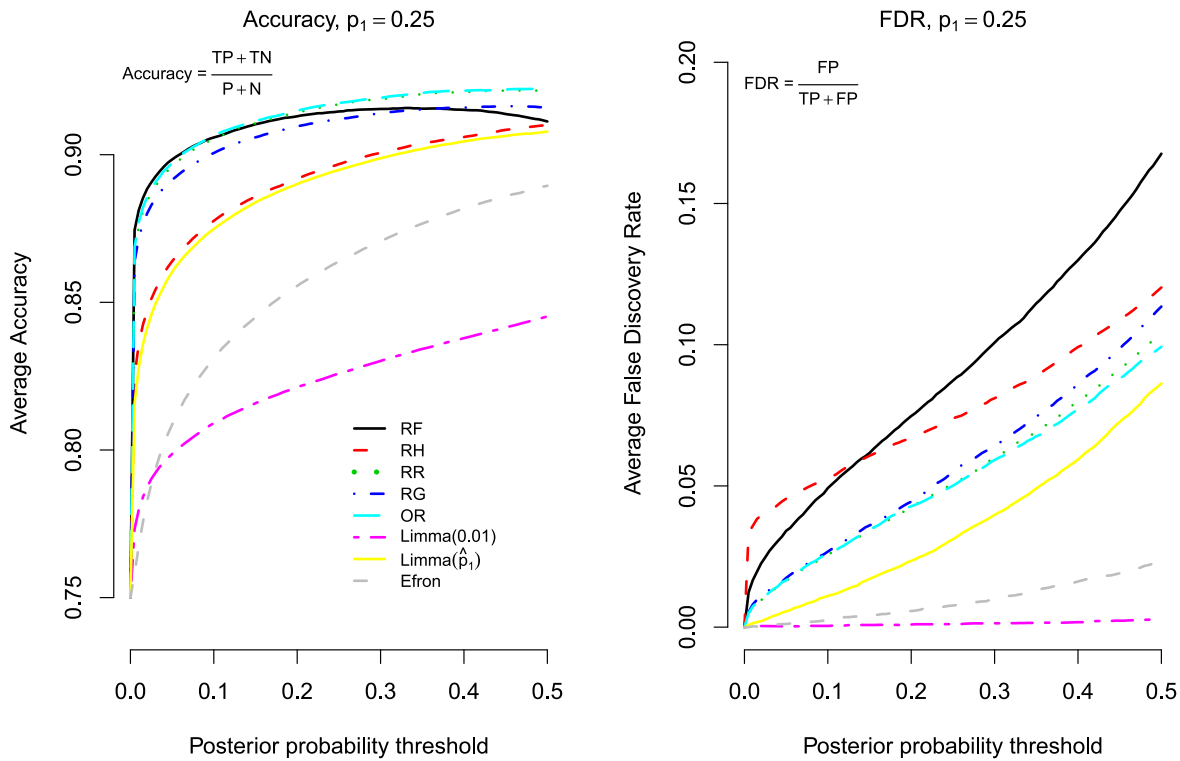}

\caption{Accuracy \textup{(left)} and false discovery rate \textup{(right)} for data
generated under the LIMMA model with
$n_1=n_2=6,$ $S=100$ samples, $G=2000$ genes, and $v_0=1$, $p_1=0.05$
probability of
non-null status, and high error variance variability ($\mathit{CV}=3.16$).}
\label{fig:LIMMA.acc.fdr.2}
\end{figure*}

Figures \ref{fig:RR.acc.fdr.1} and \ref{fig:LIMMA.acc.fdr.2} demonstrate
that when the data are generated under the LEMMA model, the RR
procedure achieves the highest level of accuracy for any posterior
probability threshold in the range [0, 0.5], and is practically the
same as the Optimal Rule. It has only a slightly higher FDR, compared
with the Optimal Rule. Note that RF has high accuracy, but very high
FDR, indicating it is too liberal and declares too many genes as
non-null.

\begin{figure*}

\includegraphics{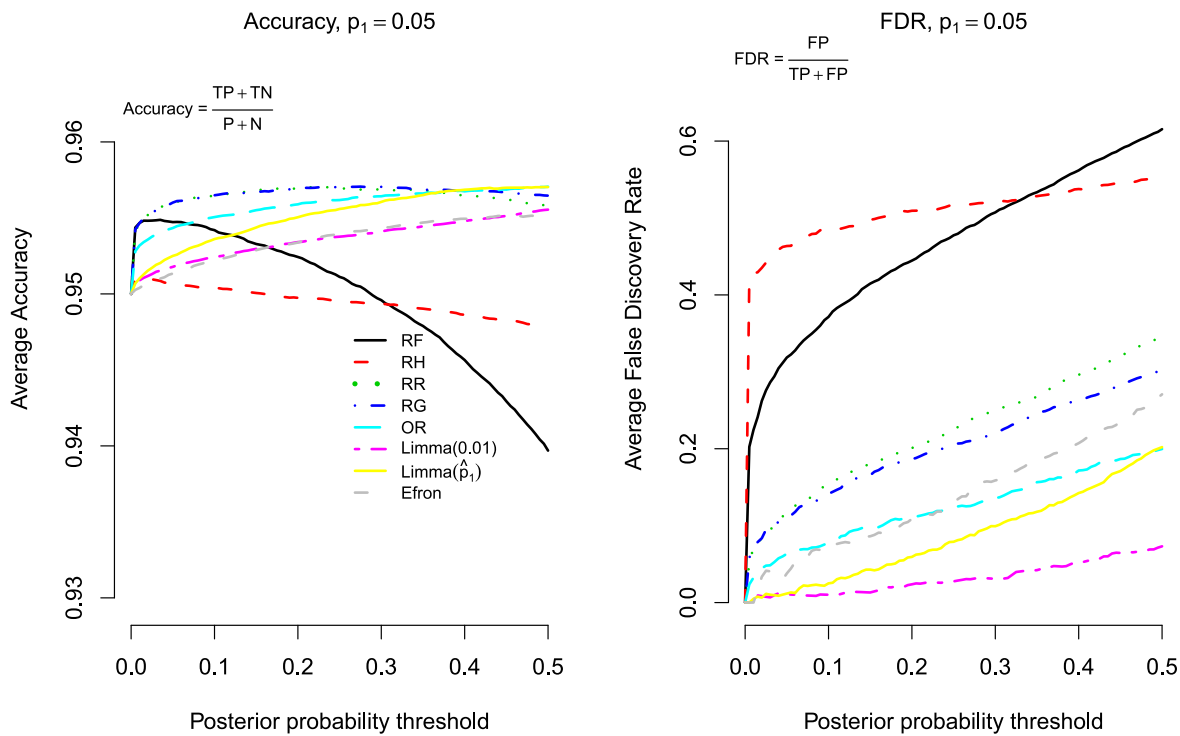}

\caption{Accuracy \textup{(left)} and false discovery rate \textup{(right)} for data
generated under the LEMMA model with
$n_1=n_2=6,$ $S=100$ samples, $G=2000$ genes, and $\psi=3$, $p_1=0.25$
probability of
non-null status, and high error variance variability ($\mathit{CV}=3.16$).}
\label{fig:RR.acc.fdr.3}
\end{figure*}

We also observe that the RR and RG procedures are quite similar, which
is an indication that the choice of the non-null variance model (either
$\sigma^2_\psi$ as in LEMMA, or $v_0\sigma^2_{\varepsilon,g}$ as
in LIMMA) does not have a significant impact on the performance. We
also notice that when the \texttt{limma} package is used with the
estimated value of $p_1$, instead of the default, the accuracy is
greatly improved, with a relatively small increase in FDR. Still, the
RR procedure (under the LEMMA data generation scheme) is clearly
superior to all other methods.

\begin{figure*}

\includegraphics{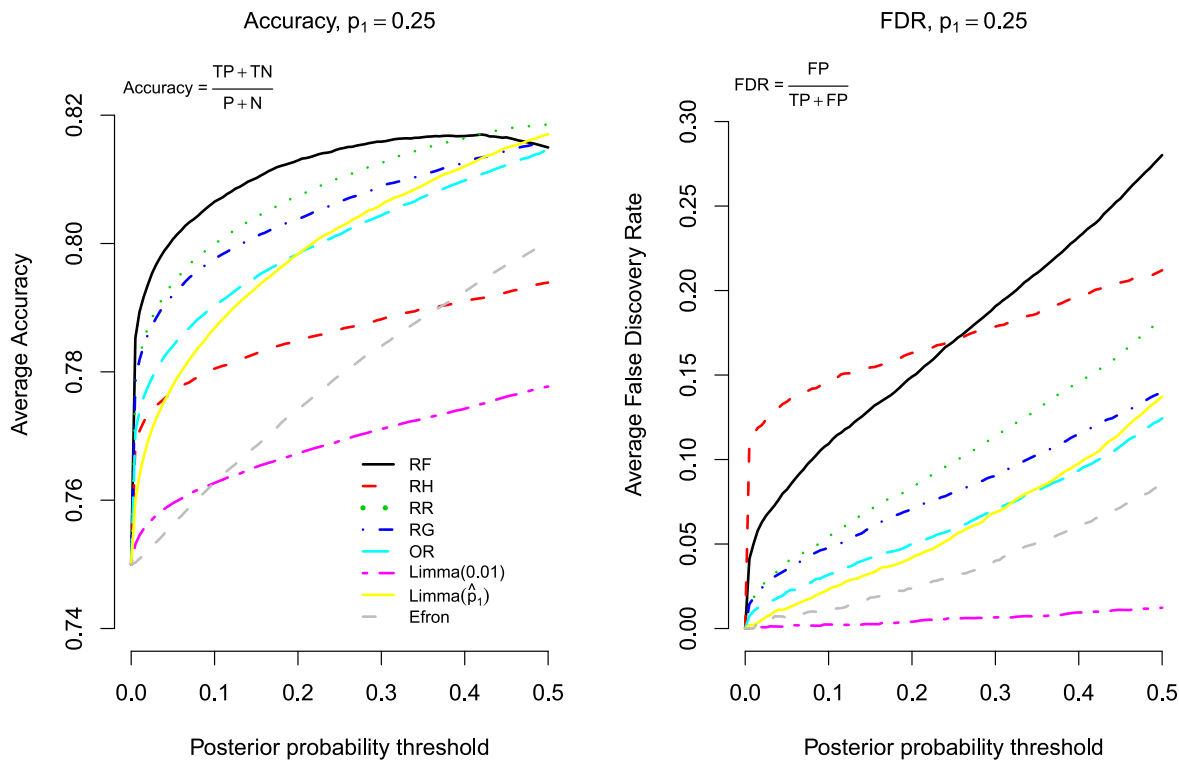}

 \caption{Accuracy \textup{(left)} and false discovery rate \textup{(right)} for data
generated under the LIMMA model with
$n_1=n_2=6,$ $S=100$ samples, $G=2000$ genes, and $v_0=1$, $p_1=0.25$
probability of
non-null status, and high error variance variability ($\mathit{CV}=3.16$).}
\label{fig:LIMMA.acc.fdr.4}
\end{figure*}

Interestingly, when the data are generated under the LIMMA model, we
get similar results---the RR procedure achieves higher accuracy, and
only a relatively small increase in false discoveries (see Figures~\ref{fig:RR.acc.fdr.3} and~\ref{fig:LIMMA.acc.fdr.4}). It is also
interesting that the \texttt{limma} procedure does not achieve the
performance of its Optimal Rule, and we believe this is due to
inaccurate estimation of $p_1,$ as demonstrated below. Note that
\texttt{lemma} uses maximum likelihood estimation for all the model
parameters, while \texttt{limma} uses ad-hoc methods to estimate $p_1$
and $v_0$. In summary, \texttt{lemma} and its RG variant are
competitive with \texttt{limma} when LIMMA is the true data generating
model, but they are clearly superior when LEMMA is the true data
generating model. Furthermore, the additional parameters
($\tau,\psi$) in the LEMMA model do not add to the computational
complexity, as the maximum likelihood estimators are obtained via a
simple, and fast EM algorithm.

To conclude this subsection, we remark that although it is possible to
compute posterior probabilities using the \texttt{limma} package (which
involves plugging in the estimates for $v_0$ and $p_1$), in practice,
inference via the \texttt{limma} package is often frequentist in nature
(using the $p$-values, computed from the $t$-statistics, returned by the
eBayes function).

\subsection{Estimation Performance}
\begin{figure*}

\includegraphics{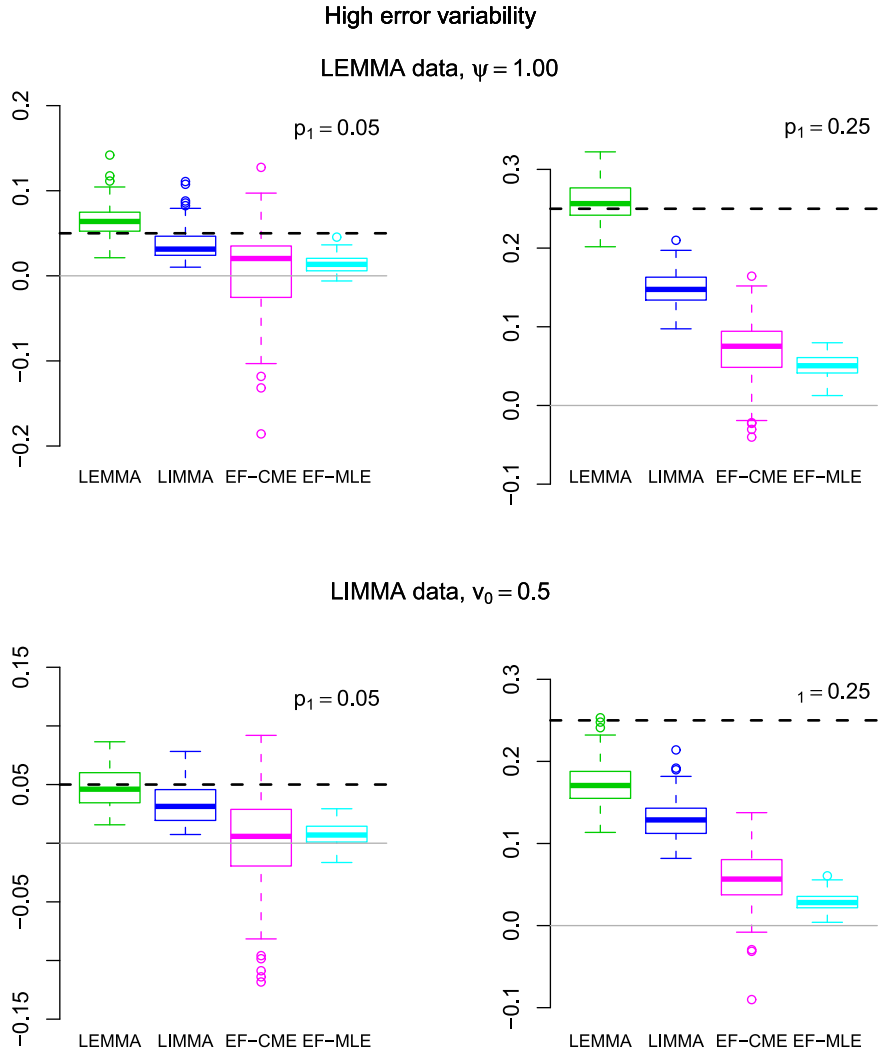}

\caption{Estimates of $p_1$ when the data are generated under the LEMMA
model (top) and under LIMMA (bottom), when the true values of $p_1$ are
0.05 \textup{(left)} and 0.25 \textup{(right}).}
\label{fig:p1.box}
\end{figure*}

We also analyzed the parameter estimation performance of the
\texttt{lemma} software, and we found it to be very accurate when the
data are generated under the LEMMA model. However, since this is not
unexpected, we chose to present a more interesting result. Recall
that both LEMMA and LIMMA require estimation of the non-null prior
probability, $p_1$. We compared the estimation of this important
parameter under those two data generation models using four estimation
methods, including \texttt{lemma}, \texttt{convest} (from the
\texttt{limma} package) and two estimation procedures available in
the \texttt{locfdr} package---denoted by EF-MLE and EF-CME.
Smyth (\citeyear{smyt2004}) argues that the mixture proportion parameter is
difficult to estimate in the model leading to the $B$-statistic, and our
simulations verify that the estimates of $p_1$ produced by the
\texttt{limma} package are significantly biased. (As noted earlier, the
\texttt{limma} package uses value of $p_1=0.01$, rather than an
estimate.) Figure \ref{fig:p1.box} shows that when $p_1=0.05$
\texttt{lemma} tends to slightly overestimate the parameter, while the
other methods tend to underestimate it. This is in agreement with the
observation that \texttt{lemma} achieves higher accuracy, and has a
slightly higher FDR. We also point out that both estimation methods
available in the \texttt{locfdr} package not only underestimate $p_1$,
but also give unreasonable (negative) estimates. The \texttt{lemma}
estimation procedure is significantly better than the other three for
higher values of $p_1$, even when the data are generated under the
LIMMA model.

\section{Examples}\label{sec6}

Using the \texttt{lemma} software, we fitted the LEMMA model to several
microarray data sets. For
illustration purposes, we provide our analysis of two publicly available,
two-channel gene expression microarray data sets that were previously
analyzed: the ApoA1 data (Callow et al., \citeyear{dataapo}) and the Colon Cancer data
(Alon et al., \citeyear{datacolonalon}).
\subsection{ApoA1 Data}
The ApoA1 experiment (Callow et al., \citeyear{dataapo}) used gene targeting in embryonic
stem cells to produce mice lacking apolipoprotein A-1, a gene known to
play a critical role in high density lipoprotein (HDL) cholesterol
levels. Originally, 5600 expressed sequence tags (EST) were
selected. In our analysis, we used the data and normalization method
provided with the \texttt{limma} R package (Smyth, \citeyear{RLIMMA}), which
consists of
5548 ESTs, from 8 control (wild type ``black six'') mice and 8
``knockout'' (lacking ApoA1) mice. Common reference RNA was obtained by
pooling RNA from the control mice, and was used to perform expression
profiling for all 16 mice. Note that the current version of the \texttt
{limma} user's guide refers to a larger data set which contains 6384
ESTs. Qualitatively speaking, using the larger data set does not yield
different results (in terms of detecting significant genes).

\begin{figure*}

\includegraphics{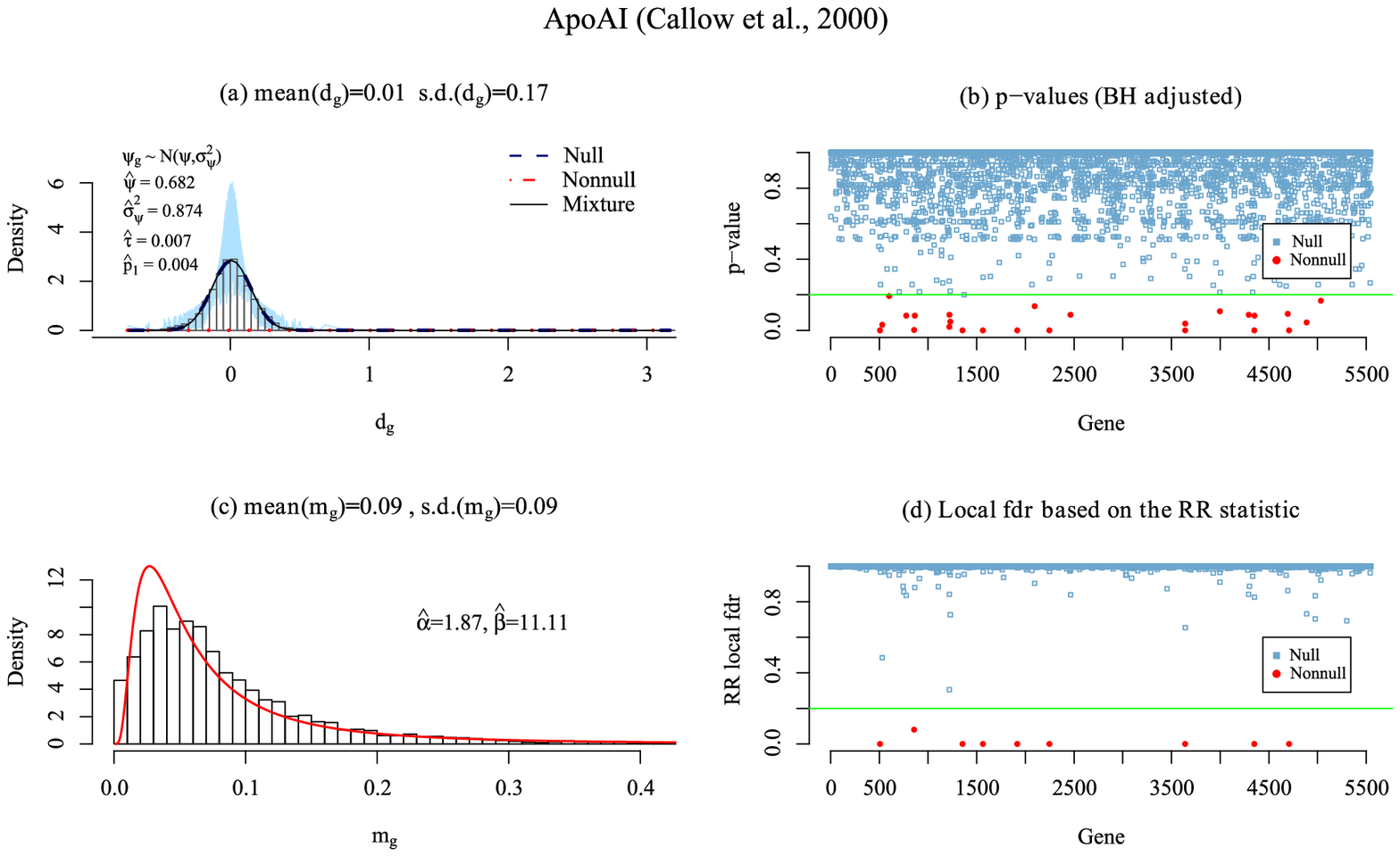}

\caption{\textup{(a)} Histogram of the 5548 $d_g$ statistics from the ApoA1 data
set and the fitted distributions. \textup{(b)} The Benjamini--Hochberg adjusted
$p$-values for all genes. Using an FDR level of 0.2, we detect 25
non-null genes. \textup{(c)} Histogram of the $m_g$ statistics and the fitted
distribution. \textup{(d)} The RR test statistics of all the genes. Using a 0.2
threshold for the posterior probability, we declare 9 genes to be non-null.}
\label{fig:apoa1.mg}
\end{figure*}

The response of interest, $y_{ijg}$, is the $\log_2$ fluorescence
ratio (with respect to the common reference) where $g$ is one of 5548
genes, $j=1,\ldots,8$ (mouse number), and $i$ is the population index
(control and knockout). Using the EM algorithm, we obtained estimates
for the parameters in our LEMMA model. Figure \ref{fig:apoa1.mg}(a)
depicts the histogram of the 5548 $d_g$ statistics. The smooth black
curve shows the fitted mixture distribution, drawn using the average
estimated error variance. The smooth blue and red curves correspond to
the average fitted distributions of the null and non-null groups,
respectively. Per-gene fitted distributions are plotted in light
colors (note that the non-null probability is very small, so only
gene-specific distributions of the null group, in light blue, can be
observed in this case). The mean-effect parameter estimates we
obtained are $\hat{\tau}=0.007$ and $\hat{\psi}=0.682$,
$\hat{\sigma}_\psi^2=0.874$.

Figure \ref{fig:apoa1.mg}(c) depicts the histogram of the $m_g$
statistics and the fitted distribution. The estimates for the shape
and scale parameters of the error variance distribution are 1.87 and
11.11, respectively. The empirical mean and variance of
$\{\tilde{\sigma}^2_{\varepsilon,g}\}$ are 0.078 and 0.004.

Using the \texttt{lemma} package, we obtained the parameter estimates,
and computed the gene-specific posterior probabilities and the
$p$-values for the hypotheses that genes are in the null group. Figure
\ref{fig:apoa1.mg}(b) depicts the Benjamini--Hochberg adjusted
$p$-values. The red, solid points represent the genes that were
declared non-null, using a (liberal) FDR threshold of 0.2. Using the
FDR criteria, we detected 25 non-null genes.

Using the posterior probabilities derived from the LEMMA (RR) model
and Efron's 0.2 threshold for local f.d.r., we detected 9 non-null genes,
including the ApoA1 gene and others that are closely related to it.
The top eight genes had local f.d.r. values of nearly zero, while the
ninth had a much higher value of 0.08. Figure \ref{fig:apoa1.mg}(d)
depicts the RR local f.d.r. statistics, and the red, solid points
represent the genes that were declared non-null using a local f.d.r.
threshold of 0.2. The top eight genes (using either the FDR or the
local f.d.r. criteria) are also identified (among others) when using the
\texttt{limma} and \texttt{locfdr R} packages, and were confirmed to
be differentially expressed in the knockout versus the control line by
an independent assay.

Interestingly, assuming no other genes are in the non-null group, the
true value of $p_1$ is 0.00144, and the estimate obtained from
\texttt{lemma} is 0.0039, while Efron's estimates using the MLE and
CME methods are $-$0.036 and $-$0.083, respectively. As we mentioned
earlier, by default the \texttt{limma R} package does not provide an
estimate for $p_1$, and uses a value of 0.01. However, using the
\texttt{convest} function, \texttt{limma} provides the estimate
$p_1=0.30$. When one uses the larger ApoA1 data set currently referred
to by the \texttt{limma} user's guide (with 6384 ESTs), the estimate
for $p_1$ is 0.134.

\subsection{Colon Cancer Data}
The data analyzed by Alon et al. (\citeyear{datacolonalon}) consists of 2000 ESTs in 40
tumor and 22 normal colon tissue samples. Of the 40 patients involved
in the study, 22 supplied both tumor and normal tissue samples. In
their analysis, Alon et al. (\citeyear{datacolonalon}) used an Affymetrix
oligonucleotide array complementary to more than 6500 human genes and
expressed sequence tags (ESTs), and a two-way clustering method to
identify families of genes and tissues based on expression patterns in
the data set. Do, M\"{u}ller and Tang (\citeyear{datacolon}) used a Bayesian mixture model to
analyze the same data set and estimated the probability of
differential expression. Using empirical Bayes methods, they obtained
a point estimate $\hat{p}_0=0.39$ and contrasted it with the posterior
marginal probability distribution of $p_0$ from the nonparametric
Bayesian model, which they fit using MCMC simulations. The empirical
Bayes estimate for $p_0$ was far out in the right tail of the
posterior distribution, which, they argued, might lead to
underestimating the posterior probability of being in the non-null
group (differentially expressed genes). They propose using posterior
expected FDR (Genovese and Wasserman, \citeyear{genovesewasserman}) thresholds to calibrate
between a desired false discovery rate and the number of significant
genes. For example, with FDR${}={}$0.2, they find 1938 non-null genes.

\begin{figure*}[t]

\includegraphics{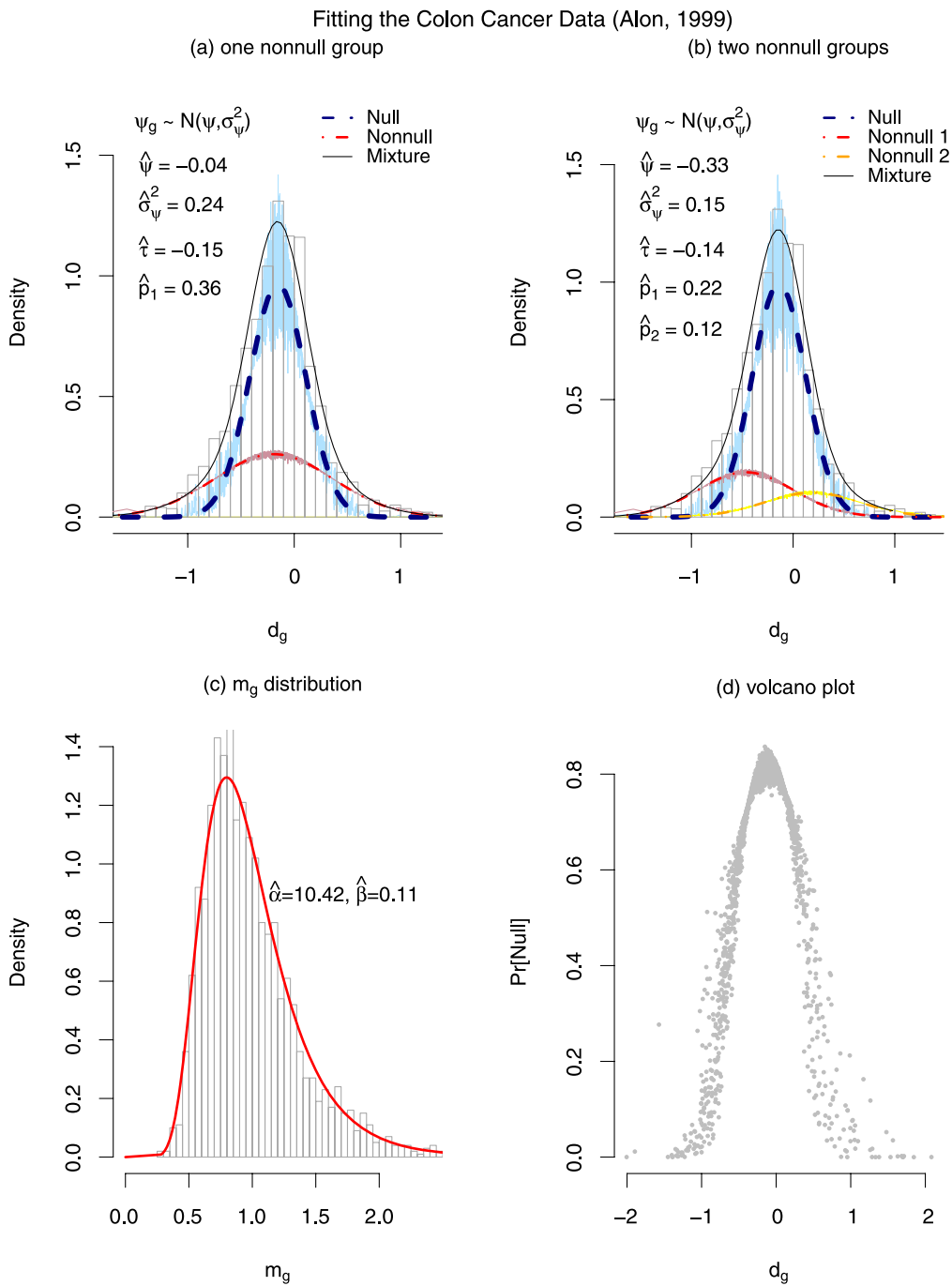}

\caption{Histograms of the 2000 $d_g$ statistics from the Alon et al.
(\protect\citeyear{datacolonalon}) data set
and the fitted distributions, assuming \textup{(a)} a two-group model, or \textup{(b)}
three-group model. \textup{(c)} Histogram of the $m_g$ statistics and the fitted
distribution. \textup{(d)} Volcano plot, showing the posterior null
probabilities by $d_g$.}
\label{fig:colon.dg}
\vspace*{-12pt}
\end{figure*}

Using \texttt{lemma} and assuming the two-group LEMMA model, we obtain
$\hat{p}_1=0.36$. According to this model, the (non-null) mean effect
of the gene-specific term is estimated by $\hat{\psi}=-0.04$ (and the
variance by $\hat{\sigma}_\psi^2=0.24$), and the fitted two-group
mixture distribution is shown in Figure \ref{fig:colon.dg}(a). The
near-zero mean of the non-null mixture component suggests that there
may be two non-null groups (over- and under-expressed groups of genes).
We fitted the three-group variant of the LEMMA model to the data, and
obtained $\hat{p}_1=0.22, \hat{p}_2=0.12,$ and $\hat{\psi}=-0.33,
\hat{\sigma}_\psi^2=0.15$ [see Figure \ref{fig:colon.dg}(b)]. In
Figure \ref{fig:colon.dg}(a) and (b) the light blue and purple curves
represent the (per gene) fitted distributions for the null and non-null
groups, respectively. The smooth black curve shows the fitted mixture
distribution, drawn using the average estimated error variance.

The three-group model allows for asymmetry in the proportions of over-
and under-expressed genes. We see no reason to assume that these
proportions should be equal. However, we find in simulations that if
they are indeed equal, our procedure estimates them accurately. We
have observed that if the true model has two non-null groups, then
estimating it assuming two modes results in an estimate of $\psi$ that
is biased toward 0 and an inflated $\hat{\sigma}^2_\psi$ (as seen in
this case), and that
this could lead to fewer true discoveries.

In this data set, the empirical mean and variance of $m_g$ are 1.00 and
0.17, respectively, with estimates $\hat{\alpha}=10.42$ and
$\hat{\beta}=0.11$. Figure \ref{fig:colon.dg}(c) shows the histogram
of the $m_g$ statistics and the fitted distribution.

The ``volcano plot'' in Figure \ref{fig:colon.dg}(d) depicts the
posterior null probability of genes based on the three-group LEMMA
model versus the $d_g$ statistics. Using the null posterior probability
threshold of 0.2, we detect 170 non-null genes, while using the FDR
method (with a threshold of 0.2) we get 155 genes. Detecting non-null
genes in a typical microarray gene expression analysis involves
setting a minimum fold-change threshold, in addition to setting the
level at which the False Discovery Rate is controlled. For instance,
requiring that $|d_g|\geq1$ and controlling the False Discovery Rate
at 0.1, we detect 61 non-null genes, all of which were detected by at
least one method in Su et al. (\citeyear{suetal2003}).\looseness=1

\section{Discussion}\label{sec7}
In the previous sections we demonstrated that our
modeling framework can lead to six different test statistics depending
on the assumptions
imposed on the gene-specific effects. Interestingly, the test
statistics associated with these models have been considered
independently in the literature in various forms, but to our
knowledge, this is the first time they have been categorized as
special cases of the same model. The LEMMA (RR) model, in which both the
non-null gene-specific effects and gene-specific variances
are modeled as random variates, leads to James--Stein-type (shrinkage)
estimation of the parameters. Specifically, the statistics derived from
the RR model enjoy
shrinkage in both the numerator and denominator of a posterior
$t$-statistic, resulting in powerful test statistics while maintaining
few false positives in our simulation studies. Using a Laplace
approximation to make the EM algorithm tractable, our approach yields
stable parameter estimates, even for the notoriously difficult
parameter~$p_1$.

Since our approach is model-based, it can be easily generalized to
other situations. For example, as stated earlier, the methods
described in this paper can be extended to deal with multiple
treatments, paired tests (one group) and multiple non-null components.
Furthermore, it is straightforward to add fixed-effect covariates to
the model. We are currently working on the next release of the
\texttt{lemma} package which will include this feature, in addition to
within-group analysis, new plotting and exporting functions, and
confidence intervals for parameter estimates. Extending the model to
handle multivariate responses is also being investigated.

\begin{appendix}
\section*{Appendix}\label{sec:app}
In this section we provide details on some of our previous derivations,
and elaborate on the case of multiple treatments.

\subsection{Empirical Bayes Estimates for $\alpha$ and $\beta$}
To obtain an estimate of the error variance in the random error case,
recall that
%
\renewcommand{\theequation}{\arabic{equation}}
\setcounter{equation}{18}
\begin{equation} \label{mgcond}
\quad m_{g}|\sigma_{\varepsilon,g}^{2} \sim\frac{\sigma_{\varepsilon
,g}^{2}}{f_{g}}\chi_{f_{g}}^{2}\equiv \operatorname{Gamma}\biggl(
\frac{f_{g}}{2},\frac{2\sigma_{\varepsilon,g}^{2}}{f_{g}}\biggr).
\end{equation}

We maximize the marginal density of $m_g$ numerically to obtain maximum
likelihood estimates of $\alpha$ and~$\beta$. Given the conditional
distribution in (\ref{mgcond}),
we find the marginal density of $m_g$ by integrating out $\sigma
^2_{\varepsilon,g}$. Specifically,
%
\begin{eqnarray}\label{mgdens.long}
f(m_g)& =&  \int_0^\infty f(m_g|\sigma^2_{\varepsilon,g})f(\sigma
^{-2}_{\varepsilon,g})\,d\sigma^{-2}_{\varepsilon,g}\nonumber\\
&=&  \int_0^\infty\biggl[ \frac{m_g^{{f_g}/{2} - 1} \exp(-{m_g f_g}/(2\sigma^2_{\varepsilon,g}))}
{\Gamma({f_g}/{2})({2\sigma^2_{\varepsilon
,g}}/{f_g})^{{f_g}/{2}}} \biggr]\nonumber\\
&&\hspace*{15pt}{}\cdot\biggl[\frac{\exp(-\sigma^{-2}_{\varepsilon,g}\beta^{-1}
)}{\Gamma(\alpha)\beta^\alpha}(\sigma^{-2}_{\varepsilon,g}
)^{\alpha-1} \biggr]\,d\sigma^{-2}_{\varepsilon,g} \nonumber\\
& =&
\frac{m_g^{{f_g}/{2}-1}({f_g}/{2})^{{f_g}/{2}}}{\Gamma({f_g}/{2})\Gamma(\alpha)\beta^\alpha}
\nonumber
\\[-8pt]
\\[-8pt]
\nonumber
&&{}\cdot \int_0^\infty(\sigma^{-2}_{\varepsilon,g})^{{f_g}/{2}+\alpha-1} \\
&&\hspace*{24pt}{}\cdot\exp\biggl[ -\sigma^{-2}_{\varepsilon,g}\biggl( \frac{m_g
f_g}{2} + \frac{1}{\beta} \biggr) \biggr] \,d\sigma^{-2}_{\varepsilon,g} \nonumber\\
&= & \frac{m_g^{{f_g}/{2}-1}({f_g}/{2})^{{f_g}/{2}}}
{\Gamma({f_g}/{2})\Gamma(\alpha)\beta^\alpha}\nonumber\\
&&{}\cdot\frac{\Gamma({f_g}/{2}+\alpha)}{({m_g f_g}/{2} + {1}/{\beta}
)^{{f_g}/{2}+\alpha}}.\nonumber
\end{eqnarray}
The final equality in (\ref{mgdens.long}) results from noting that the
integral in the third equality is proportional to a $\textrm
{Gamma}(f_g/2+\alpha, [\beta^{-1} + m_g f_g/2]^{-1})$ density. We
maximize $\sum_g{\log(f(m_g))}$ with respect to $\alpha$ and $\beta$ to
obtain the empirical Bayes estimates $\hat{\alpha}$ and $\hat{\beta}$.

The joint distribution of $m_g$ and $\sigma_{\varepsilon,g}$ is given by
\begin{eqnarray*}
f( m_{g},\sigma_{\varepsilon,g}^{-2})& =&
m_{g}^{f_{g}/2-1}f_{g}^{f_{g}/2}( \sigma_{\varepsilon,g}^{-2})
^{\alpha
-1+f_{g}/2}\\
&&\cdot{}\exp\biggl\{ -\sigma_{\varepsilon,g}^{-2}\biggl[ \frac
{m_{g}}{2}f_{g}+\frac{%
1}{\beta}\biggr] \biggr\} \\
&&{}\Big/{\biggl(\Gamma\biggl( \frac{f_{g}}{2}\biggr)
2^{f_{g}/2}\Gamma( \alpha) \beta^{\alpha}\biggr)}.
\end{eqnarray*}
So, conditional on $m_{g}$,
\[
\sigma_{\varepsilon,g}^{-2}\sim \operatorname{Gamma}\bigl(\alpha+f_{g}/2,(
m_{g}f_{g}/2+1/\beta) ^{-1}\bigr).
\]
Hence, the conditional expectation is
\begin{eqnarray*}
&&E( \sigma_{\varepsilon,g}^{2}|m_{g})\\
&&\quad=\frac
{f_{g}/2}{f_{g}/2+\alpha-1}
m_{g}+\frac{\alpha+ 1 }{f_{g}/2+\alpha-1}\cdot\frac{1}{(\alpha+1)
\beta} \\
&&\quad\approx\frac{f_{g}/2}{f_{g}/2+\alpha-1}m_{g}+\frac{\alpha+ 1 }{
f_{g}/2+\alpha-1}\bar{m},
\end{eqnarray*}
and the conditional mode is
\begin{eqnarray*}
&&\mathit{Mode}( \sigma_{\varepsilon,g}^{2}|m_{g}) \\
&&\quad=\frac
{f_{g}/2}{f_{g}/2+\alpha
+1}m_{g}+\frac{\alpha+1}{f_{g}/2+\alpha+1}\cdot\frac{1}{(\alpha+1)
\beta}
\\
&&\quad\approx\frac{f_{g}/2}{f_{g}/2+\alpha+1}m_{g}+\frac{\alpha+1 }{
f_{g}/2+\alpha+1}\bar{m}.
\end{eqnarray*}
Note that using the approximation of the mode, $\bar{m} \approx[(\alpha
+1)\beta]^{-1},$ in both the posterior mean and posterior mode yields a
shrinkage-estimator form. Equivalently, we could replace $(\alpha+1)$
with $(\alpha-1)$ in the conditional expectation and the conditional
mode, and obtain shrinkage toward the sample mean of $\{m_g\}$.

\subsection{Maximum Likelihood Estimation of $\phi$}
Recall that in the RR method we use the Laplace approximation (\ref
{eq:laplace}), hence, the (approximate) complete likelihood is
%
\begin{eqnarray}\label{comp.lik.lap.long}
&&\tilde{L}_{C}(\phi)\nonumber\\
&&\quad\propto \prod_{g=1}^G L(b_g, d_g; \tilde{\sigma
}_{g}^{2})\nonumber\\
&&\quad= \prod_{g=1}^G \int L(b_g;p_{1}) L( d_{g}|b_g;\psi_{g},\tilde{\sigma
}_{g}^{2}
)\nonumber\\
&&\qquad\hspace*{27pt}{}\cdot f( \psi_{g}|b_g)\, d\psi_{g}
\nonumber
\\[-8pt]
\\[-8pt]
\nonumber
&&\quad= \prod_{g=1}^G \biggl[
p_{1}^{b_{g}}( 1-p_{1}) ^{1-b_{g}}( 2\pi\tilde{\sigma
}_{g}^{2}) ^{-{1}/{2}}\\
&&\hspace*{17pt}\qquad{}\cdot(2 \pi\sigma_{\psi
}^{2}) ^{-{b_{g}}/{2}}\exp\biggl\{ -\frac{1-b_{g}}{2\tilde
{\sigma}_{g}^{2}}( d_{g}-\tau) ^{2}\biggr\} \nonumber\\
&&\hspace*{17pt}\qquad{}\cdot\int\exp\biggl\{ -\frac{b_{g}}{2\tilde{\sigma}_{g}^{2}}
( d_{g}-\tau-\psi_g) ^{2}\nonumber\\
&&\hspace*{17pt}\qquad{}-\frac{b_{g}}{2 \sigma_{\psi
}^{2}}( \psi_{g}-\psi) ^{2}\biggr\}\,d\psi_{g}
\biggr] ,\nonumber
\end{eqnarray}
with log-likelihood
%
\begin{eqnarray}\label{comp.log.lik.lap}
\ell(\phi)
&\propto&\sum_{g=1}^{G} [ ( 1-b_{g}) \log(
1-p_{1})
+b_{g}\log( p_{1}) ]\nonumber \\
&&{}-\sum_{g=1}^{G}\biggl[ \frac{b_{g}}{2}\log(2 \pi\sigma_{\psi
}^{2}) +\frac{1}{2}\log( 2\pi\tilde{\sigma}_{g}^{2})
\biggr] \nonumber\\
&&{}-\frac{1}{2}\sum_{g=1}^{G}b_{g}\log\biggl((2 \pi)
^{-1}\biggl( \frac{1}{
\tilde{\sigma}_{g}^{2}}+\frac{1}{\sigma_{\psi}^{2}}\biggr) \biggr)\\
&&{}-\sum_{g=1}^{G}( 1-b_{g}) \frac{( d_{g}-\tau)
^{2}}{2\tilde{\sigma}_{g}^{2}}\nonumber\\
&&{}-\sum_{g=1}^{G}\frac{b_{g}}{2}\biggl[ \frac{1}{\sigma_{\psi}^{2}+
\tilde{\sigma}_{g}^{2}}( d_{g}-\tau-\psi)
^{2}\biggr].\nonumber
\end{eqnarray}

The estimates (\ref{eq:p1.update})--(\ref{eq:sigma2psi.update}) are
obtained by maximizing the log-likelihood with respect to the
parameters, $\phi$.

Although the Laplace approximation is not necessary in the RF and RH
models, note that the complete likelihoods and log-likelihoods for the
these models are identical to equations (\ref{comp.lik.lap.long}) and
(\ref{comp.log.lik.lap}), with $\tilde{\sigma}_{g}^{2}$ replaced by
$\hat{\sigma}_{g}^2$ and $\hat{\sigma}_{\varepsilon}^2$ (as defined in
Section \ref{sec:mods}), respectively.

\subsection{Multiple Treatments}
In the general case we assume $t\geq2$ treatments $i=1,2,\ldots,t$
assigned to $t$ groups of $n_{1,g},n_{2,g},\ldots,n_{t,g}$ subjects
indexed by $j_{1}=1,\ldots,n_{1,g}, \ldots, j_{t}=1,\ldots,\break n_{t,g}$, and we
use the model defined by (\ref{eq:linear.model}) and (\ref
{eq:error.distribution}). Here, we impose a standard (fixed effect) constraint
\[
\sum_{i=1}^{t}\psi_{ig}=0.
\]
The distributions for the gene-specific effects in the
multiple-treatment case are assumed to follow
a normal distribution,
\begin{eqnarray*}
\bolds{\psi}_{g}\sim\mbox{i.i.d.}~N_{t}\bigl(\bolds{\psi} ,\sigma
_{\psi}^{2}( \mathbf{I}_{t}-\mathbf{\bar{J}}_{t}) \bigr),
\end{eqnarray*}
where $\mathbf{I}_t-\mathbf{\bar{J}}_{t}$ is the $t \times t$ centering
matrix, $\bolds{\psi}$ is a $t$-dimensional vector, and $\sigma
_{\psi}^{2}$ is a scalar.
The test statistic $m_g$ is defined as
\[
m_g=\sum_{i=1}^t\sum_{j=1}^{n_{ig}}(y_{ijg}-\bar{y}_{i\cdot g})^2/f_g ,
\]
where $f_g=n_{1g}+\cdots+n_{tg}-t$, and we use $m_g$ as before, to
estimate $\alpha$ and $\beta$.

To estimate the rest of the parameters in the LEMMA model, we use the
($t-1$-dimensional vector) test statistics
\[
\mathbf{d}_{g}=\mathbf{H\bar{Y}}_{g},
\]
where
\begin{eqnarray*}
\mathbf{H} &=&\left[\matrix{
1/\sqrt{2} & -1/\sqrt{2} & 0  \cr
1/\sqrt{6} & 1/\sqrt{6} & -2/\sqrt{6}  \cr
\vdots & \vdots & \vdots& \cr
1/\sqrt{t( t-1) } & 1/\sqrt{t( t-1) } & 1/\sqrt{t( t-1) }}\right.\\
&&\hspace*{80pt}{}\left.\matrix{\cdots& 0\cr
\cdots& 0\cr
& \vdots\cr
\cdots& -(
t-1) /\sqrt{t( t-1)}
}\right] ,\\
\mathbf{\bar{Y}}_{g} &=&( \bar{Y}_{1\cdot g},\bar{Y}_{2\cdot g
},\ldots,\bar{Y}_{t\cdot g}) ^{\prime
}.
\end{eqnarray*}

Derivations similar to the ones we used to obtain the estimates in
Section \ref{sec:estimation} lead to the same estimate for $p_1$ and to
the following estimates, analogous to (\ref{eq:tau.update}) and~(\ref
{eq:psi.update}):
\begin{eqnarray*}
&&\bigl( \mathbf{H}{\bolds{\tau}}^{(m+1)}\bigr) ^{\prime}\\
&&\quad=\Biggl[
\sum_{g=1}^{G}p^{(m)}_{0,g} \mathbf{d}_{g}^{\prime}\bolds{\Lambda
}_{0}^{-1}\Biggr] %
\Biggl[ \sum_{g=1}^{G}p^{(m)}_{0,g}\bolds{\Lambda}_{0}^{-1}\Biggr]
^{-1},
\\
&&\bigl( \mathbf{H\bolds{\psi}}^{(m+1)}\bigr)^{\prime}\\
&&\quad=\Biggl[
\sum_{g=1}^{G}p^{(m)}_{1,g}\bigl( \mathbf{d}_{g}-\mathbf{H}\bolds
{\tau}^{(m+1)} \bigr) ^{\prime}( \bolds{\Lambda}_{0}+\bolds
{\Lambda}_{A}) ^{-1}\Biggr] \\
&&\qquad{}\cdot\Biggl[
\sum_{g=1}^{G}p^{(m)}_{1,g}( \bolds{\Lambda}_{0}+\bolds{\Lambda
}_{A})
^{-1}\Biggr] ^{-1},
\end{eqnarray*}
where
\begin{eqnarray*}
\bolds{\Lambda}_{A} &=&\sigma_{\psi}^{2(m)}\mathbf{I}_{t-1},\\
\bolds{\Lambda}_{0} &=&\tilde{\sigma}_{\varepsilon,g}^{2} \mathbf{H}
[ \operatorname{diag}_{i}( 1/n_{ig}) ] \mathbf{H}^{\prime}.
\end{eqnarray*}

The update for $\tilde{\sigma}_{\psi}^{2}$ is the solution to the equation,
\begin{eqnarray*}
&&\sum _{g=1}^{G}p^{(m)}_{1,g}\cdot \operatorname{tr}\bigl( ( \bolds{\Lambda
}_{0}+\bolds{\Lambda}_{A}) ^{-1}\bigr)\\
&&\quad=\sum
_{g=1}^{G}p^{(m)}_{1,g}\bigl( \mathbf{d}_{g}-\mathbf{H}\bolds{\xi}^{(m+1)}
\bigr) ^{\prime}( \bolds{\Lambda}_{0}+\bolds{\Lambda
}_{A}) ^{-2}\\
&&\qquad{}\cdot\bigl( \mathbf{d}_{g}-\mathbf{H}\bolds{\xi
}^{(m+1)}\bigr),
\end{eqnarray*}
where $\bolds{\xi}^{(m+1)}=\bolds{\tau}^{(m+1)}+\bolds
{\psi}^{(m+1)}$.

The likelihood ratio test statistic has a similar form as (\ref
{eq:likelihood.ratio}),
\begin{eqnarray*}
\frac{L_{0,g}}{L_{1,g}} &=&\vert\mathbf{I}-\bolds{\Lambda
}_{g}\vert
^{-1/2}
\exp\biggl\{ -\frac{1}{2} \bolds{\Gamma} ^{\prime}(\bolds
{\Lambda}_{0}^{-1}\bolds{\Lambda}_{g}^{-1}) \bolds{\Gamma}
\biggr\}\\
&&{}\cdot\exp\biggl\{ -\frac{1}{2}\tilde{\sigma}_{\psi}^{-2}( \mathbf
{H}\hat{\bolds{\psi}} ) ^{\prime}(\mathbf{H}\hat
{\bolds{\psi}}) \biggr\},
\end{eqnarray*}
where
\begin{eqnarray*}
\bolds{\Gamma} &=& [ \bolds{\Lambda}_{g}( \mathbf
{d}_{g}-\mathbf{H\hat{\bolds{\tau}} }) +( \mathbf
{I}-\bolds{\Lambda}_{g}) ( \mathbf{H}\hat
{\bolds{\psi} }) ], \\
\bolds{\Lambda}_{g} &=&( \bolds{\Lambda}_{A}+\bolds{\Lambda
}_{0}) ^{-1} \bolds{\Lambda}_{A} , \\
\mathbf{I}-\bolds{\Lambda}_{g} &=&( \bolds{\Lambda}_{A}+\bolds
{\Lambda}_{0}) ^{-1}\bolds{\Lambda}_{0}.
\end{eqnarray*}
\end{appendix}

\section*{Acknowledgments}

We would like to thank Yoav Benjamini, Daniel Yekutieli, Peng Liu and
Gene Hwang, and particularly Gordon Smyth who carefully examined
earlier versions of this paper and provided us with a number of
important comments and suggestions.
We are also grateful to two anonymous referees for their useful
comments and suggestions.
James Booth supported in part by NSF Grant DMS 0805865.
Martin T. Wells supported in part by NIH Grant R01-GM083606-01.

\end{document}